\documentclass[11pt]{article}
\usepackage{moriond,epsfig}

\def\be{\begin{equation}}
\def\ee{\end{equation}}
\def\bea{\begin{eqnarray}}
\def\eea{\end{eqnarray}}

\newcommand{\ov}[1]{\overline{#1}}
\newcommand{\Bbar}{\,\overline{\!B}}

\newcommand{\gqbtf}{\ensuremath{\Gamma (\Bbar{}_q(t) \rightarrow f )}}
\newcommand{\gqtfb}{\ensuremath{\Gamma (B_q(t) \rightarrow \ov{f} )}}

\begin{document}
\begin{flushright}
CERN-PH-TH/2012-111
\end{flushright}

\vspace*{4cm}
\title{Theoretical update of $B$-Mixing and Lifetimes}

\author{ Alexander Lenz }

\address{CERN - Theory Divison, PH-TH, Case C01600, CH-1211 Geneva 23}

\maketitle\abstracts{
 We review the current status of theoretical predictions 
 for  mixing quantities and lifetimes in the $B$-sector. 
In particular, due to the first non-zero measurement of 
the decay rate difference in the neutral $B_s$-system,
$\Delta \Gamma_s/ \Gamma_s  = 17.6 \% \pm 2.9 \%$ by the LHCb collaboration 
and very precise  data for $\tau_{B_s}$ from TeVatron and LHCb 
our theoretical tools can now be rigorously tested and it turns 
out that the Heavy Quark Expansion works in the $B$-system to an accuracy 
of at least $30\%$ for quantities like $\Gamma_{12}$, which is most sensitive
to hypothetical violations of  quark hadron duality.
This issue that gave rise in the past to numerous theoretical papers, 
has now been settled experimentally. Further data will even allow to 
shrink this bound. For total inclusive quantities like lifetimes
the compliance is even more astonishing: 
$\tau_{B_s}^{\rm LHCb}/ \tau_{B_d}^{\rm HFAG} = 1.001 \pm 0.014$ is in 
perfect agreement with the theory expectation of 
$\tau_{B_s}/\tau_{B_d} = 0.996 ... 1.000$.
\\
Despite the fact that the new data show no deviations from the standard model expectations,
there is still some sizable room for new physics effects. Model-independent search strategies
for these effects are presented with an emphasis on the interconnection 
with many different observables that have to be taken into account. In that respect a special 
emphasis is given to the large value of the di-muon asymmetry measured by the D0 collaboration.}

\section{Introduction}

\begin{center}
\begin{figure}[h]
\begin{center}
\includegraphics[width=0.69\textwidth,angle=0]{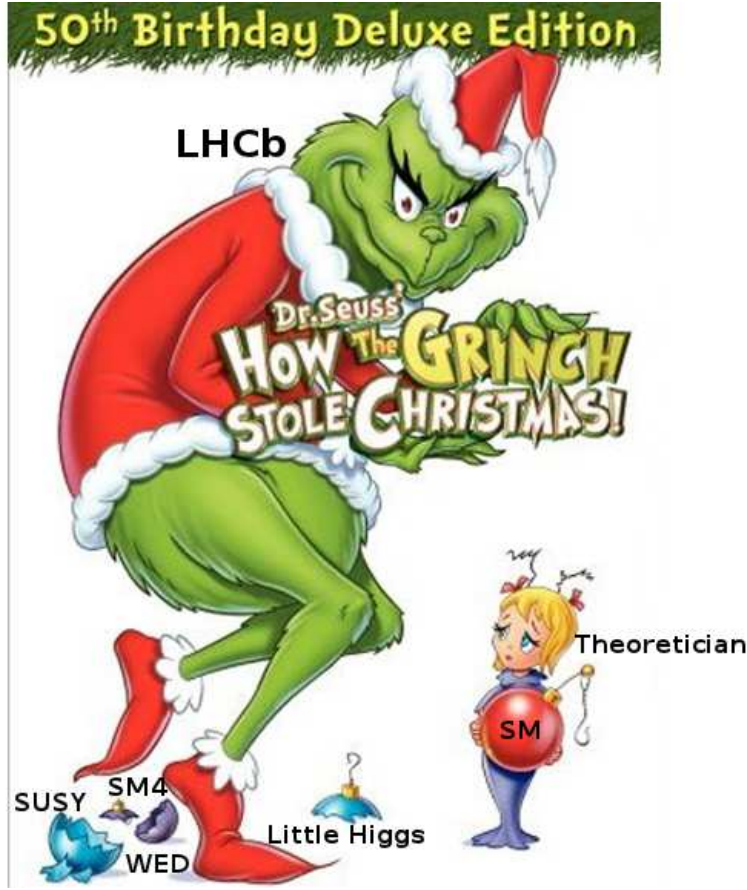}
\caption{How the new LHCb results might appear at first sight. Previous measurements at TeVatron raised 
big expectations in large new physics effects, which were unfortunately  not confirmed by more precise
measurements at LHCb. It is, however, important to note that we gained many theoretical insights by the first
measurement of $\Delta \Gamma_s$ and by other very precise measurements. This will be elaborated in 
this paper.}
\label{grinch}
\end{center}
\end{figure}
\end{center}

Recently\footnote{Before the {\it Lepton Photon} Conference in August 2011.}
several hints for new physics effects in the flavour sector, see e.g.
\cite{Lenz:2010gu,Lunghi:2010gv,Buras:2010xj,Ligeti:2010ia}, triggered a lot of interest.
In particular there were several independent experimental indications
\cite{Abazov:2011ry,Abazov:2011yk,Abazov:2010hj,Abazov:2010hv,Abazov:2008af,Aaltonen:2007he} 
for a large CP violating phase in $B_s$-mixing, which would be in 
clear contradiction to the tiny standard model (SM) expectation 
\cite{Lenz:2011ti,Lenz:2006hd,Beneke:2003az,Ciuchini:2003ww,Laplace:2002ik}.
\\
Unfortunately the huge expectations in spectacular new physics effects were not confirmed
by precise measurements at LHCb or by follow-up measurements at the TeVatron,
e.g. \cite{LHCb:2012,LHCb:2011aa,CDF:2011af,Lenz:2012az}.
\\
The current disappointment about this situation can be nicely visualized with Fig.(\ref{grinch}).
\footnote{ A recent review on the implications of this new experimental results on several 
new physics models can be found in \cite{Buras:2012ts}.}

One of the main aims of this talk is to emphasize the fact that we now have
for the first time a measurement of a non-zero value of the decay rate difference in
the neutral $B_s$-system, $\Delta \Gamma_s$\cite{LHCb:2012}. 
This quantity raised a lot of interest in the last thirty-five years and several
open questions related to the theoretical determination of $\Delta \Gamma_s$ 
have now been answered experimentally. Another important point is the fact that
there is still some ample space for new physics effects in $B$-mixing.
To distinguish possible new physics effects from hadronic uncertainties, however, 
a considerably higher theoretical precision for questions like penguin pollution 
is now necessary, compared to what would have been needed in the case of a huge mixing phase.
\\
In Section 2 we give a short introduction to the mixing formalism and to its experimental evidence.
We also briefly touch the calculational tools, in particular the Heavy Quark Expansion (HQE). 
In Section 3 historic attacks to the HQE and their experimental resolution are described. 
Therefore we discuss theoretical arguments in favour of the HQE as well as the missing charm puzzle,
the lifetimes of the $\Lambda_b$-baryon and the $B_s$-meson, 
the decay rate differences $\Delta \Gamma_d$ and $\Delta \Gamma_s$ 
and semi leptonic CP asymmetries. This chapter contains the main results: the proof of the applicability 
of the HQE to determine $\Gamma_{12}$.
Model independent investigations of possible new physics effects in mixing are presented in Section 4, while
Section 5 concentrates on new physics contributions and its constraints to the absorptive
part of the mixing diagram, $\Gamma_{12}$. Such effects were discussed in order to explain the
large central value of the di-muon asymmetry measured by the D0 collaboration. Since it turned out 
experimentally that new physics effects in $B$-mixing are not huge, 
investigations of the remaining space for  effects beyond the standard model require 
now a much better theoretical control of e.g. penguin contributions, this is discussed in Section 6.
Finally we conclude.

\section{Mixing Formalism}

\subsection{Basic mixing formulae}

The time evolution of a decaying particle
$ B(t) = \exp \left[ -i m_B t - \Gamma_B/2 t \right] $ with mass $m_B$ and lifetime 
$\tau_B = 1/ \Gamma_B$
can be written as a differential equation. In the case of a two-state system (e.g. the neutral 
meson $B_q = (\bar{b}q)$ and its anti-particle $\bar{B_q}=(b \bar{q})$) one expects a diagonal
mass matrix $\hat{M}^q$ and a diagonal decay rate matrix $ \hat{\Gamma}^q$:
\begin{equation}
i \frac{d}{dt}
\left(
\begin{array}{c}
| B_q(t) \rangle \\ | \bar{B}_q (t) \rangle
\end{array}
\right)
=
\left( \hat{M}^q - \frac{i}{2} \hat{\Gamma}^q \right)
\left(
\begin{array}{c}
| B_q(t) \rangle \\ | \bar{B}_q (t) \rangle
\end{array}
\right) \; .
\end{equation}
It is known since a long time (e.g. \cite{Gaillard:1974hs})
that due to weak interactions
transitions like { $B_{d,s} \to \bar{B}_{d,s}$} are possible
in the neutral $B$-system. They are triggered by the so-called
{\it box diagrams}.
  \centerline{\includegraphics[width=0.8\textwidth,angle = 0]{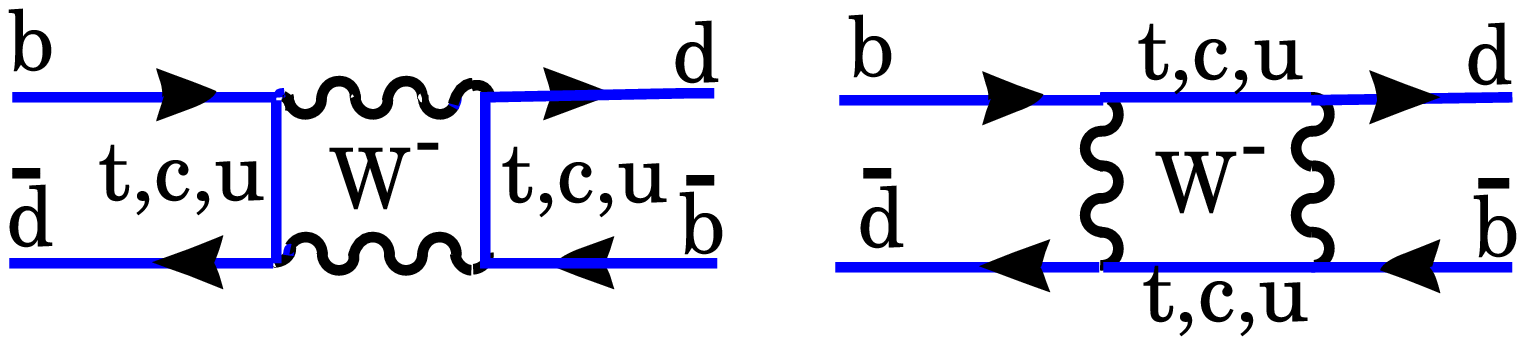}}
The box diagrams lead 
to off-diagonal elements in the mass       matrix  $\hat{M}_q$ 
(denoted by $M_{12}^q$) 
and 
to off-diagonal elements in the decay rate matrix  $\hat{\Gamma}_q$ 
(denoted by $\Gamma_{12}^q$). 
$M_{12}^q$ corresponds to the part of the box-diagram, where the internal quarks are off-shell
(e.g. the internal top-quark), 
while $\Gamma_{12}^q$ denotes the part where the internal quarks
are on-shell (in the diagrams shown above this can only happen for the up- and charm-quark).
Both $M_{12}^q$ and $\Gamma_{12}^q$ can be complex due to their dependence on
CKM-elements.
\\
Diagonalization of $\hat{M}^q$ and $ \hat{\Gamma}^q$ gives mass eigenstates
$B_{q,H}$ (H=heavy) and $B_{q,L}$ (L=light) with the masses $M_H^q,\,M_L^q$ and the decay rates
$\Gamma_H^q,\,\Gamma_L^q$, see e.g.\cite{GellMann:1955jx}:
\begin{eqnarray}
B_{q,H} := p \; B_q - q \; \bar{B}_q 
&, &
B_{q,L} := p \; B_q + q \; \bar{B}_q
\, \, \, \mbox{  with} \, \, \,  |p|^{2} + |q|^{2}  =  1 \;.
\end{eqnarray}
In the limit of no CP violation in mixing, the heavy mass eigenstate is CP-odd and the light one
CP-even.
Theoretical calculations of the box diagrams ($M_{12}^q$ and $\Gamma_{12}^q$) can be related
to three observables, expressed in terms of
$  |M_{12}^q|$, $|\Gamma_{12}^q|$ and the relative phase 
$  \phi_q = \mbox{arg}( -M_{12}^q/\Gamma_{12}^q)$, which describes CP-violation in mixing; 
the individual phases of $M_{12}^q$ and $\Gamma_{12}^q$ are unphysical:
\begin{itemize}
\item \underline{\bf Mass difference:} 
      \\
      \\
      The mass difference between the heavy and the light mass eigenstates of the neutral 
      $B_q$ meson is given by the modulus of the dispersive part of the box diagram.
       \begin{equation}
       \Delta M_q := M_H^q - M_L^q  = 
        2 { |M_{12}^q|} \left( 1 -
        {  \frac{1}{8} \frac{|\Gamma_{12}^q|^2}{|M_{12}^q|^2} \sin^2 \phi_q + ...}\right) \; .
       \label{DeltaM}
       \end{equation}
       In the standard model $|\Gamma_{12}^q|/|M_{12}^q|$ is of the order of 5 per mille
       \cite{Lenz:2011ti,Lenz:2006hd} for the $B_q$-system. Therefore the corrections in Eq.(\ref{DeltaM}) 
       are completely negligible. 
       Since $M_{12}^q$ describes the virtual part of the box diagram, it is
       sensitive to heavy internal particles like the top-quark, but also to
       hypothetical new states like SUSY-particles, KK-states, ... . So, the mass difference
       is expected to be very sensitive to new physics effects.
\item \underline{\bf Decay rate difference:}
      \\
      \\
      The decay rate  difference between the heavy and the light mass eigenstates of the neutral 
      $B_q$ meson is given by the modulus of the absorptive part of the box diagram.
      \begin{equation} 
       \Delta \Gamma_q := \Gamma_L^q - \Gamma_H^q = 
        2 { |\Gamma_{12}^q| \cos  \phi_q }
        \left( 1 +
        {  \frac{1}{8} \frac{|\Gamma_{12}^q|^2}{|M_{12}^q|^2} \sin^2 \phi_q 
        + ...}\right) \; .
        \end{equation}
        In the standard model $\phi_s$ is close to zero \cite{Lenz:2011ti,Lenz:2006hd}, 
        so the cosine gives to a good approximation a value of one.
        Since $\Gamma_{12}^q$ describes the on-shell part of the box diagram it is
        sensitive to light (with a total mass below $M_B$) internal particles, like the up- and
        charm-quark. As we do not expect sizable new physics contributions to CKM-favoured tree-level
        decays like $b \to c \bar c s$, it seems reasonable to assume almost no (i.e. below 
        the hadronic uncertainties) new physics effects in $\Gamma_{12}^s$.
        Therefore the only way for new physics to affect $\Delta \Gamma_s$ is the phase $\phi_s$.
        In the standard model $\phi_s$ is close to zero, so new physics can only lower
        $\Delta \Gamma_s$ compared to its standard model value \cite{Grossman:1996era}.
        We will challenge, however, the assumption that $\Gamma_{12}^q$ can not be affected 
        by new physics in Section 5. 
\item \underline{\bf Flavour specific /semi leptonic CP asymmetries:}
      \\ \\
      A flavour specific decay $B_q \to f$ is defined by the property
      \begin{itemize}
      \item The decays { $\bar{B}_q \to f$} and { $B_q \to \bar{f}$} are forbidden.
      \end{itemize}
      To simplify the analysis of untagged decays one often demands in addition 
      that no direct CP violation occurs in the decay, i.e.
      { $|\langle f | B_q \rangle| =| \langle\bar{f} | \bar{B}_q\rangle|$}.
      Examples of such a decay are  $B_s \to D_s^- \pi^+$ or the semi leptonic 
      decay { $B_q \to X l \nu$} - hence also the name semi leptonic CP asymmetry.
      The asymmetry is defined as
      \begin{eqnarray}
       a_{sl}^q & \equiv & a_{fs}^q = \frac{\gqbtf - \gqtfb}{\gqbtf + \gqtfb}
      = 
      - 2 \left( \left| \frac{q}{p}\right| - 1 \right) 
      \nonumber
      \\
      &= & \left| \frac{\Gamma_{12}^q}{M_{12}^q}\right| \sin \phi_q
      =  { \mbox{Im} \frac{\Gamma_{12}^q}{M_{12}^q}}
      =  \frac{\Delta \Gamma_q}{\Delta M_q} \tan \phi_q \; .
      \end{eqnarray}
      In the standard model these asymmetries are suppressed by the small values of 
      $|\Gamma_{12}^q|/|M_{12}^q|$ 
      and $\phi_q$. 
      Measuring a sizable value for the semi leptonic CP asymmetries would be a clear indication
      for new physics, provided that we have a precise control over the theoretical framework.
      Theoretical arguments in favour of this control and the experimental proof of this control, 
      which is now available,  will be discussed in detail below.
\end{itemize}
The phenomenon of particle-antiparticle mixing is a macroscopic quantum effect, which is by now 
well established in several systems of neutral mesons.
\begin{itemize}
\item[1955] $K^0$-system: Mixing in the neutral $K$-system was theoretically developed in 
            1955 by Gell-Mann and Pais \cite{GellMann:1955jx}. Based on that framework
            the phenomenon of {\it regeneration} was predicted in the same year by Pais and 
            Piccioni  \cite{Pais:1955sm}. Experimentally regeneration was confirmed in 
            1960 \cite{Muller:1960ph}. A huge lifetime difference between the two neutral
            $K$-mesons was established already in 1956 \cite{Lande:1956pf}.
\item[1986] $B_d$-system: Mixing in the $B_d$-system was found 1986 by UA1 at CERN \cite{Albajar:1986it} 
            (UA1 attributed the result however to $B_s$ mixing)
            and 1987 by ARGUS at DESY\cite{Albrecht:1987dr}. The large result for
            the mass difference $\Delta M_d$ can be seen as the first clear hint for an 
            (at that time) unexpected large value of the top quark mass\cite{Ellis:1987mm}
            \footnote{To avoid a very large value of the top quark mass, also different new physics
            scenarios were investigated, in particular a scenario with a heavy fourth generation
            of fermions and a top quark mass of the order of 50 GeV, see e.g. \cite{Tanimoto:1987wv}.}.
            For the decay rate difference currently only upper bounds are available, 
            see \cite{Higuchi:2012kx} for
            the most recent and most precise bound.
\item[2006/12] $B_s$-system: The large mass difference in the $B_s$-system was established by the CDF
            collaboration at TeVatron \cite{Abulencia:2006ze}. In 2012 the LHCb Collaboration
            measured for the first time a non-vanishing value of the decay rate difference in the 
            $B_s$-system \cite{LHCb:2012}.
\item[2007] $D^0$-system: Here we have several experimental evidences (BaBar, Belle, Cleo, CDF, E791, E831)
            for values of
            $\Delta \Gamma / \Gamma$ and $\Delta M / \Gamma$ at the per cent level,
            but we still do not have a single measurement with a statistical significance
            of more than five standard deviations. The combination
            of the data\cite{Asner:2010qj} shows, however,  unambiguously that D mesons mix. 
\end{itemize}
A more detailed discussion of the experimental evidence for mixing can be found e.g. in \cite{gersabeck}.
\\
Because of several reasons, mixing observables seem to be very well suited for the 
search of hypothetical new physics effects.
First, these observables are experimentally established - as summarized above.
Next, in the standard model the box diagrams are strongly suppressed, because of being 
fourth order in the weak interaction. So even small new physics effects might give comparable
and therefore detectable  contributions. Finally, hadronic effects are well under control
in the $B$ mixing observables. This will be elaborated below.

\subsection{Theoretical determination of $M_{12}$}
Calculating the virtual part of the box diagrams with internal up-, charm- and top-quark and
using the unitarity of the 3x3 CKM matrix one finds that the by far dominant contribution is 
given by the top-quark. It reads
        \begin{eqnarray}
        M_{12}^q & = & \frac{G_F^2}{12 \pi^2} 
          (V_{tq}^* V_{tb})^2 M_W^2 S_0(x_t)
          { B_{B_q} f_{B_q}^2  M_{B_q}} \hat{\eta }_B \; .
        \label{M12SM}
        \end{eqnarray}
The evaluation of the 1-loop box-diagram gives the so-called Inami-Lim function $S_0(x_t)$
\cite{Inami:1980fz} with $x_t = m_t^2/M_W^2$.
Perturbative QCD corrections to the box-diagrams are denoted by { $\hat{\eta }_B$}
\cite{Buras:1990fn}, they turned out to be  ample.
In the end one also has to determine the non-perturbative matrix element of 
a four quark operator $Q$ (the four external legs of the box diagram) switched in between
the states of the physical $B$ and $\bar B$ mesons. For historical reasons this 
matrix element is parameterized in terms of a bag parameter $B_{B_q}$ and a decay constant
$f_{B_q}$:
\begin{eqnarray}
\langle \bar{B_q}|
Q
|B_q \rangle
& = & 
\frac{8}{3} { B_{B_q} f_{B_q}^2  M_{B_q}^2} \; .
\label{Bag}
\\
Q & = & \bar{q}_\alpha \gamma_\mu (1- \gamma_5) b_\alpha
\cdot
\bar{q}_\beta \gamma^\mu (1-\gamma_5) b_\beta
\label{Q}
\end{eqnarray}
$\alpha$ and $\beta$ denote colour indices.
The bag parameter and the decay constant have to be determined with non-perturbative methods like
lattice QCD or QCD sum rules.
\\
Formally one has performed in Eq.(\ref{M12SM}) an operator product expansion (OPE), by integrating
out the heavy $W$-boson and the heavy top-quark. Corrections to that expansion are expected
to be of the order $m_b^2/M_W^2$, i.e. smaller than four per mille and thus far below the 
hadronic uncertainties.
\\
Comparing experiment (HFAG\cite{Asner:2010qj}) with theory \cite{Lenz:2011ti} one sees
a very nice agreement, but also the fact that the theoretical
error is considerably larger than the experimental one.
The theory error is currently dominated by the hadronic parameters.
        \begin{eqnarray}
         \Delta M_d^{\rm SM} = 0.543 \pm 0.091 \; \mbox{ps}^{-1}, & & \Delta M_d = 0.507 \pm 0.004 \; \mbox{ps}^{-1}
        \label{DeltaMd} \\
        &&  \mbox{(ALEPH, CDF, D0, DELPHI, L3, }
        \nonumber
        \\
        && \mbox{OPAL, BABAR, BELLE, ARGUS, CLEO)},
        \nonumber
        \\
        \Delta M_s^{\rm SM} = 17.30 \pm 2.6  \; \mbox{ps}^{-1}, & & \Delta M_s = 17.69 \pm 0.08  \; \mbox{ps}^{-1}
        \label{DeltaMs}\\
        & & \mbox{(CDF, D0, LHCb)} .
        \nonumber
        \end{eqnarray}
For the theory prediction in Eq.(\ref{DeltaMd}) and Eq.(\ref{DeltaMs}) taken from \cite{Lenz:2011ti}  
a very conservative average \cite{Lenz:2010gu} of lattice determinations for the matrix element 
in Eq.(\ref{Bag}) was taken as an input 
\footnote{For a comparison of different ways to average the lattice results, see e.g. the
discussion in \cite{Wingate:2011fb}.}.
In particular 
\begin{equation}
f_{B_s} = 231 \pm 15 \; \mbox{MeV}
\end{equation}
was used for the decay constant\footnote{For more recent averages, see e.g.\cite{Lenz:2012az,Laiho:2009eu}.}.
Recently several new evaluations of the decay constant were 
performed\cite{Dimopoulos:2011gx,McNeile:2011ng,Bazavov:2011aa}.
With this new numbers one obtains:
\begin{equation}
\begin{array}{l|c|c}
\mbox{Reference} & f_{B_s} & \Delta M_s
\\
\hline \hline
\mbox{HPQCD (1110.4510)}  & 225 \pm 4 \; \mbox{MeV} & 16.4 \pm 1.0 \; \mbox{ps}^{-1}
\\
\mbox{Fermilab/MILC (1112.3051)} & 242 \pm 10 \; \mbox{MeV} & 19.0 \pm 1.8 \; \mbox{ps}^{-1}
\\
\mbox{CKMfitter new average (1203.0238)} & 229 \pm 6 \; \mbox{MeV} &  17.1 \pm 1.7 \; \mbox{ps}^{-1}
\end{array}
\label{newDeltaM}
\end{equation}
For the first two results for $\Delta M_s$  in Eq.(\ref{newDeltaM}) only the value and the error 
of the decay constant was changed compared to Eq.(\ref{DeltaMs}). For the third number
also a new average for the bag parameter was taken into account.
In view of the relative large difference of the central values of the decay constants (17 MeV)
determined in \cite{McNeile:2011ng,Bazavov:2011aa} we prefer to stay currently with our conservative 
prediction given in Eq.(\ref{DeltaMs}).
\\
Here a further reduction of the theoretical errors and a future reduction of the difference between 
distinct evaluations will be very desirable, because the mass differences give 
important bounds on fits \cite{CKMfits} of the unitarity triangle and also to searches for new physics effects.

\subsection{Theoretical determination of $\Gamma_{12}$}

The theoretical determination of $\Gamma_{12}^q$ is much more complicated than the one of
$M_{12}^q$, sketched in the previous subsection.
$\Gamma_{12}^q$ is sensitive to real intermediate states like the up- and the charm-quark.
Thus one can not integrate out at once all particles inside the loop of the box diagram,
one has to follow instead a two step procedure:
\begin{enumerate}
\item OPE I:  Integrate out the heavy W boson. This is similar to the case of $M_{12}^q$.
      Again, corrections to that approximation are expected to be at most at the per mille level.
\item OPE II: Peform an expansion in inverse powers of the heavy quark mass, the 
      heavy quark expansion (HQE)
\cite{Khoze:1983yp,Shifman:1984wx,Chay:1990da,Bigi:1992su,Bigi:1993fe,Blok:1993va,Manohar:1993qn}.
\end{enumerate}
Then the expression for $\Gamma_{12}^q$ can be written in the following form:
        \begin{eqnarray}
        \Gamma_{12}^q \! \! & \! \!= \! \!&\! \!
        \left(\frac{\Lambda}{m_b}\right)^3 \! \! \left( \Gamma_3^{(0)} +
                                                   \frac{\alpha_s}{4 \pi} { \Gamma_3^{(1)}} +... \right) +
        \left(\frac{\Lambda}{m_b}\right)^4 \! \! \left( \Gamma_4^{(0)} + ... \right)  +
        \left(\frac{\Lambda}{m_b}\right)^5 \! \! \left( {\Gamma_5^{(0)}} + ... \right)  + ... \; .
         \end{eqnarray}
$\Lambda$ is a mass scale, which is expected to be of the order of $\Lambda_{QCD}$. Its precise
value has, however, to be determined by explicit calculation. As will be shown below, for the $B_s$-system
the expansion parameter $\Lambda / m_b$ is a little smaller than 1/5.
The $\Gamma_i^{(j)}$'s are products of perturbatively calculable Wilson coefficients $C_k$ and
non-perturbative matrix elements of operators with a mass dimension equal or larger than six.
Such a matrix element is also parameterized in terms of the decay constant $f_B$ and a 
bag parameter $B_k$, i.e.
$\Gamma_i^{(j)}\propto f_B^2 \sum C_k B_k$.
\\
First estimates of $\Gamma_3^{(0)}$ were made starting from 1977 on, see e.g. 
\cite{Ellis:1977uk,Hagelin:1981zk,Franco:1981ea,Chau:1982da,Buras:1984pq,Khoze:1986fa}.
At that time no effective Hamiltonian and no non-perturbative determinations
of the decay constant and the bag parameters were available. Since 1977 clearly the
determination of $\Gamma_3^{(0)}$ improved a lot. 
The subleading $1/m_b$-corrections $\Gamma_4^{(0)} $ were determined for the
first time in 1996 \cite{Beneke:1996gn} for $B_s$ mesons, they turned out to be quite large.
The NLO-QCD corrections $\Gamma_3^{(1)}$ for the leading CKM-structure in $\Delta \Gamma_s$ were calculated 
already in 1998 \cite{Beneke:1998sy}.
In 2003 this result was extended to the subleading CKM-structure by two independent
groups \cite{Beneke:2003az,Ciuchini:2003ww}. These subleading terms give, however, the dominant contribution
to the semi leptonic CP asymmetries and $\Delta \Gamma_d$, while they are a small, 
but non-negligible contribution to the decay rate difference $\Delta\Gamma_s$. 
In 2006 \cite{Lenz:2006hd} several theoretical improvements were incorporated in the theoretical determination
of $\Gamma_{12}^s$, like an use of different quark mass schemes or summing up large logarithms of the
form $m_c^2/m_b^2 \ln m_c^2/m_b^2$ to all orders (following \cite{Beneke:2002rj}) as well as a change of 
the operator basis.
Calculating $\Gamma_{12}$ to the order $\Lambda^4/m_b^4$ one gets
\begin{equation}
\Gamma_{12}^q = C   \langle Q           \rangle_q +  
              C_S \langle Q_S         \rangle_q + 
      \tilde{C}_S \langle \tilde{Q}_s \rangle_q + 
             \sum \limits_{i=1,2,3} \left( C_{R_i} \langle R_i  \rangle_q +
                            \tilde{C}_{R_i} \langle \tilde{R}_i \rangle_q \right) \; ,
\end{equation}
with Wilson coefficients $C$, $C_S$, $\tilde{C}_S$ and $C_R$ and matrix elements of the
operator $Q$ appearing in $M_{12}$ (defined in Eq.(\ref{Q})) and of  two new operators 
\begin{eqnarray}
Q_S & = &\bar{q}_\alpha (1+ \gamma_5) b_\alpha
         \cdot
         \bar{q}_\beta  (1+\gamma_5) b_\beta \; ,
\label{QS}
\\
\tilde{Q}_S & = &\bar{q}_\alpha (1+ \gamma_5) b_\beta
         \cdot
         \bar{q}_\beta  (1+\gamma_5) b_\alpha \; .
\label{tildeQS}
\end{eqnarray}
$R_i$ is an acronym for different operators of dimension seven, appearing in 
$\Gamma_4$, which can be found e.g. in \cite{Lenz:2006hd}.
In \cite{Beneke:1996gn} it was noted that $Q$, $Q_S$, $\tilde{Q}_S$ are not 
independent, they are related via
\begin{equation}
R_0 = Q_S + \alpha_1 \tilde{Q}_S + \frac{1}{2} Q \; .
\label{R0}
\end{equation}
The coefficients have the form $\alpha_i = 1 + {\cal O}(\alpha_s(m_b))$ 
and $R_0$ is of order $1/m_b$.
So, to leading order in $1/m_b$ the quantity $\Gamma_{12}^q$ can be expressed in
terms of two independent operators, e.g. $(Q, Q_S)$, denoted as the {\it old basis} or 
$(Q, \tilde{Q}_S)$, denoted as the {\it new basis}. 
\begin{eqnarray}
\mbox{old basis:} && \! \! \Gamma_{12}^q =  C^{\rm old}   \langle Q           \rangle_q +  
              C_S^{\rm old} \langle Q_S         \rangle_q + 
          C_{R_0}^{\rm old}  \langle R_0         \rangle_q +
            \! \! \! \! \! \!  \sum \limits_{i=1,2,3} \left( C_{R_i} \langle R_i  \rangle_q+
                            \tilde{C}_{R_i} \langle \tilde{R}_i \rangle_q \right) .
\label{old}
\\
\mbox{new basis:} && \! \! \Gamma_{12}^q =  C^{\rm new}   \langle Q           \rangle_q +  
      \tilde{C}_S^{\rm new} \langle \tilde{Q}_s \rangle_q + 
          C_{R_0}^{\rm new}  \langle R_0         \rangle_q +
             \! \! \! \! \! \! 
                \sum \limits_{i=1,2,3} \left( C_{R_i} \langle R_i  \rangle_q+
                            \tilde{C}_{R_i} \langle \tilde{R}_i \rangle_q \right)  .
\label{new}
\end{eqnarray}
In principle both expressions are absolutely equivalent.
But, since the bag parameters of the different operators are not known with 
the same precision the choice of the basis has a sizable effect on the resulting 
accuracy in the determination of $\Gamma_{12}^q$.
The best known matrix element is the one of the operator $Q$. This operator also 
arises in the mass difference $\Delta M_q$ and there are many non-perturbative 
determinations on the market. For the matrix element of $Q_S$ or $\tilde{Q}_s$ much 
less calculations are available, while for the matrix element of $R_0$ 
only vacuum insertion approximation is used - this corresponds to an uncertainty 
of $50 \%$!
Looking now at the sizes of the coefficients in Eq.(\ref{old}) and Eq.(\ref{new}) one finds\cite{Lenz:2006hd} 
\begin{itemize}
\item $C_{R_0}^{\rm new}<C_{R_0}^{\rm old}$: In the new basis the coefficient of the
      operator $R_0$ is considerably smaller than in the old basis - 
      this favours the new basis, because the matrix element of $R_0$ has a very large uncertainty.
\item In the old basis $C_S^{\rm old}$ is the dominant coefficient, in the new basis
      the coefficient $ C^{\rm new}$ of the best known operator is dominant -
      this favours the new basis.
\item In the old basis, there is an accidental strong numerical cancellation 
      in the coefficient $C^{\rm old}$, which can lead to an overestimate of the
      theoretical uncertainty in the old basis. 
\item In the new basis the by far dominant contribution to the theory prediction 
      for $\Delta \Gamma_q / \Delta M_q$ is absolutely free of non-perturbative 
      uncertainties. This is not the case in the old basis, which is therefore 
      disfavoured.
\end{itemize}
Thus we will use the new basis until very precise lattice values for all arising operators
and the $\alpha_s /m_b$-corrections will be available.
In 2007 also the Wilson coefficients of the $1/m_b^2$ corrections, i.e. 
$\Gamma_5^{(0)}$ have been determined \cite{Badin:2007bv}. Their size seems to be
reasonably small. Since for some of the arising dimension eight operators even
vacuum insertion approximation is not applicable, we do not include these terms 
in our numerics.
\section{HQE under attack}
Performing the HQE implicitly assumes that the sum over all possible exclusive
final states, which is necessary to determine the total lifetime, is equal to 
the sum over all possible final state quarks. This assumption is called quark
hadron duality, see e.g.\cite{Poggio:1975af} or \cite{Bigi:2001ys} for the 
application to decays of heavy hadrons.
For total decay rates one obtains in the framework of the HQE
        \begin{eqnarray}
        \Gamma & = & \Gamma_0 
        + \left(\frac{\Lambda}{m_b}\right)^2 \Gamma_2
        + \left(\frac{\Lambda}{m_b}\right)^3 \Gamma_3
        + \left(\frac{\Lambda}{m_b}\right)^4 \Gamma_4
        + ... \; .
        \label{HQEall}
        \end{eqnarray}
The leading term $\Gamma_0$ describes the decay of the free b-quark and is 
therefore the same for all $b$-hadrons. It depends on the fifth power of 
the b-quark mass, so depending on what quark mass scheme one is using, 
one gets very different results.
Therefore it is advantageous to look at ratios of lifetimes of different
$b$-hadrons. For different mesons this ratio starts with contributions
of the order $\Lambda^3 / m_b^3$ (for baryon vs. meson lifetime there are 
already corrections of the order of $\Lambda^2 / m_b^2$), which look very similar
to the contributions of $\Gamma_{12}^q$ \footnote{A more detailed comparison 
of the theoretical determination of lifetime differences of different b-hadrons 
and of $\Gamma_{12}^q$ is given e.g. in \cite{Lenz:2008xt}.}.
Comparing lifetime predictions with measurements provides a test
of the assumption of quark hadron duality. Although one should keep in mind that
for $\Gamma_{12}^q$ one needs a stronger assumption, because less intermediate
states (only the ones that are common to $B_q$ and $\bar{B}_q$) contribute to the sum.
To draw some definite conclusion on $\Gamma_{12}^q$ also a direct determination of e.g.
$\Delta \Gamma_q$ is mandatory, which is now available for $B_s$ mesons\cite{LHCb:2012}!
\\
In the last 20 years regularly some discrepancies between experiment and theoretical
predictions assuming quark hadron duality have been arising, which always 
triggered a lot of theoretical interest.
Below I will discuss the following topics:
\begin{itemize}
\item In the mid 90ies the {\it missing charm puzzle} received a lot of attention, see e.g
      \cite{Lenz:2000kv} for a mini-review. The measured average number of charm-quarks
      per $b$-decay turned out to be lower than theoretically predicted:
      $n_c^{\rm Exp.} < n_c^{\rm SM}$. This discrepancy was also related to a 
      mismatch of the experimental and theoretical values for the semi leptonic 
      branching ratio. A possible solution was the violation of quark hadron 
      duality, see e.g. \cite{Bigi:1993fm,Falk:1994hm}.
\item From the beginning of the 90ies the values for the lifetime of the $\Lambda_b$
      baryon was measured to be much shorter, than the lifetime of the $B_d$ mesons,
      while theory predicted them to be of similar size.
      This was also attributed to a failure of local duality in 
      inclusive non leptonic heavy flavour decays, see e.g. \cite{Altarelli:1996gt}.
\item Before 2003 the values of  $\tau_{B_s}/ \tau_{B_d} $ were smaller than $0.95$.
      Here theory predicts almost exactly one. This would also hint to a violation
      of quark hadron duality or to new physics effects in the lifetimes.
\item In 2010 \cite{Abazov:2010hj,Abazov:2010hv} and 2011 \cite{Abazov:2011yk}
      the D0 collaboration announced an unexpected large value of the 
      di-muon asymmetry. It turned out that assuming new physics in $M_{12}^q$ is
      not sufficient to explain the large central value, one also needs large 
      deviations of $\Gamma_{12}^q$ from its standard model value, either via a
      violation of quark hadron duality or via new physics effects, see e.g. \cite{Bobeth:2011st}.
\end{itemize}
In the following we will show that the above puzzles disappeared or that the solution 
via a violation of quark hadron duality is not viable anymore.

\subsection{Theory arguments in favour of the HQE}
In this section I will give some theoretical arguments in favour of the validity of the 
HQE. These arguments clearly represent no proof - for that an exact solution of 
QCD would have to be compared with the HQE, which is beyond the scope of the current paper.
    \begin{enumerate}
    \item A pragmatic starting point is to calculate corrections within the HQE
          in all possible ``directions'', in order to test the convergence of the expansion.
          Splitting up the contributions to $\Delta \Gamma_s$
          in a leading term (no perturbative QCD corrections, 
          vacuum insertion approximation for the bag parameters and leading term in 
          the HQE) and subleading corrections due to the use of lattice values 
          for the bag parameters, due to perturbative QCD corrections and due to 
          subleading HQE corrections
          one gets \cite{Lenz:2011zz}:
        \begin{eqnarray}
        \Delta \Gamma_s & = &  \Delta \Gamma_s^0 \left(
         1 + \delta^{\rm Lattice} + \delta^{\rm QCD} + \delta^{\rm HQE} \right)
        \nonumber
        \\
        & = & 0.142 \; \mbox{ps}^{-1} \left( 1 -0.14 - 0.06 - 0.19 \right) \; .
        \label{convergence}
        \end{eqnarray}
        Explicit calculation shows that the corrections are at most $19\%$, which 
        seems to be a reasonable value and indicates no breakdown of the theory. 
        In particular the expansion parameter of the HQE
        is now determined to be $0.19 \approx 1/5$. So no signal for a breakdown 
        of the HQE is found by an explicit 
        calculation\footnote{This statement is of course much more solid than 
        statements based on phase space arguments and simple power counting, 
        see e.g. the discussion in\cite{Lenz:2011zz}.}.
    \item As mentioned above (see \cite{Lenz:2008xt} for more details) the theoretical 
         expressions for lifetime ratios of mesons resemble the expression for $\Gamma_{12}^q$.
        \begin{eqnarray}
        \frac{\tau_{B_1}}{\tau_{B_2}} - 1 \! \! & \! \!= \! \!&\! \!
        \left(\frac{\Lambda}{m_b}\right)^3 \! \! \left( \Gamma_3^{(0)} +
                                                   \frac{\alpha_s}{4 \pi} { \Gamma_3^{(1)}} +... \right) +
        \left(\frac{\Lambda}{m_b}\right)^4 \! \! \left( \Gamma_4^{(0)} + ... \right)  +
        \left(\frac{\Lambda}{m_b}\right)^5 \! \! \left( {\Gamma_5^{(0)}} + ... \right)  + ... \; .
        \nonumber
        \\
        \end{eqnarray}
        Moreover lifetimes are expected to be insensitive to new physics effects \footnote{Sizable new 
        physics effect in lifetimes correspond to sizable new physics effects in the dominant decay
        modes, which have not been seen experimentally.}.
        Comparing theory and experiment for $\tau (B^+) / \tau (B_d)$ one gets a very nice 
        agreement \cite{Lenz:2011ti}.
        Again, no sign for a breakdown of the OPE. 
        \\
        The precision of this comparison is however strongly limited by the corresponding
        hadronic parameters. The latest available lattice evaluation of the arising matrix elements
        stems from 2001 \cite{Becirevic:2001fy}. Here clearly an update would be very helpful.
  \item The inclusive approach to calculate $\Delta \Gamma_s$ described above, can also be 
        compared to the exclusive approach,
        where one tries to estimate all exclusive decay channels that contribute to $\Gamma_{12}^s$.
        The seminal paper of Aleksan et. al from 1993 \cite{Aleksan:1993qp} was recently 
        updated \cite{Chua:2011er} and the new results agree well with the predictions using the HQE.
        Again, an argument in favour of the validity of the HQE.   
    \end{enumerate}
    We have summarized here several theoretical hints for the applicability of the HQE. These hints
    represent of course not a proof. In the next section we show how the above listed problems of the
    HQE predictions have been resolved experimentally. 
\subsection{The missing charm puzzle}
\label{missing}
The theoretical numbers for $n_c$ are more than ten years old
\cite{nc_theory}, so a reanalysis with updated input parameters seems to be 
desirable. Using a preliminary
theoretical result \cite{nc_neu} with input parameters from 2012
and comparing them with the latest value from BaBar in 2006 \cite{Aubert:2006mp}
one finds perfect agreement.
\begin{eqnarray}
n_c^{\rm 2006 BaBar} & = & 1.208^{+0.058}_{-0.055} \; .
\\
n_c^{\rm SM}       & = & 1.20 \pm 0.04 \; .
\end{eqnarray}
So there is no hint for a violation of quark hadron duality from the charm counting
anymore.
Nevertheless it would be very desirable to have updated experimental numbers
for $n_c$ from e.g. the full data sets of the B-factories, because this could shrink 
the allowed region for hypothetical new physics effects contributing to $\Gamma_{12}^q$, 
see below.
\subsection{The $\Lambda_b$-lifetime}
The experimental number\cite{Asner:2010qj} for the $\Lambda_b$-lifetime changed 
quite dramatically in the last ten
years
\begin{eqnarray}
\mbox{HFAG '03} && \tau_{\Lambda_b} = 1.229 \pm 0.080 \; \mbox{ps}^{-1} \; .
\\
\mbox{HFAG '12} && \tau_{\Lambda_b} = 1.425 \pm 0.032 \; \mbox{ps}^{-1} \; .
\end{eqnarray}
This is a shift of the central value of about $2.5 \sigma$. Part of this change is due to the 
fact that old measurements were only done with semi leptonic $\Lambda_b$ decays, while
for newer measurements also non leptonic decays became available.
The newer values can be easily explained by the HQE, e.g. \cite{Tarantino:2007nf}, 
while the old low values  were very problematic. So again, the indications for a 
failure of the HQE have vanished.
\\
Looking a little closer, one sees however that there are still some issues that have to be settled.
Experimentally there seems to be a discrepancy of a about 2 $\sigma$ between lifetime measurements
of D0 \cite{D0_Lambdab} and CDF \cite{Aaltonen:2010pj} - both measure the non leptonic decay channel.
To settle this issue we are eagerly waiting for new LHC results.
\\
But there are also some open theoretical issues, in particular the correct value of the 
non-perturbative matrix elements of the four-quark operators of the $\Lambda_b$-baryon.
Due to HQET only two different matrix elements (instead of four) arise. A widely used
parameterization is 
      \begin{equation}
      \frac{1}{2 m_{\Lambda_b}} 
      \langle \Lambda_b |
      \bar{b}_L \gamma_\mu q_L \cdot \bar{q}_L \gamma^\mu b_L |
      \Lambda_b \rangle =: - \frac{f_B^2 m_B}{48} r \; .
      \end{equation}
The second matrix element can also be expressed in terms of the parameter $r$, so in the
end $\tau_{\Lambda_b} / \tau_{B_d}-1 $ is proportional to $r$.
Unfortunately we do not have a lattice evaluation of $r$, only an exploratory study from 1999 
\cite{DiPierro:1999tb}, whose numerical result should be taken with a lot of caution.
In the literature values of $r$ range from 0.2 (QCD sum rules \cite{Colangelo:1996ta})
over 1.2 (the exploratory lattice study \cite{DiPierro:1999tb}) up to $6.2$ (QCD sum rules
\cite{Huang:1999xj} - these results have met however a lot of scepticism in the sum rule community).
Rosner related in 1996 \cite{Rosner:1996fy} the value of $r$ to mass differences of $b$-hadrons and obtained
values between 0.9 and 1.8. Using his results and the recent PDG values \cite{Nakamura:2010zzi} of the
$b$-hadron masses we get
\begin{eqnarray}
r & \approx & \frac{4}{3} \frac{m_{\Sigma_b^*}^2 - m_{\Sigma_b}^2}{m_{B^*}^2 - m_B^2} = 0.68 \pm 0.10 \;,
\end{eqnarray} 
which points towards smaller values of $r$ and consequently to larger values of the 
$\Lambda_b$ lifetime. Here clearly new lattice determinations are mandatory to improve the theory 
prediction.

\subsection{The $B_s$-lifetime}

For quite some time the $B_s$ lifetime was also considerably below the $B_d$ lifetime. Just recently
several new results for $\Gamma_s = 1/ \tau_{B_s}$ were obtained by 
performing the angular analysis \cite{Dunietz:2000cr} in the decay $B_s \to \psi \phi$.
LHCb \cite{LHCb:2012}, CDF \cite{CDF:2012} and D0 \cite{Abazov:2011ry} obtained from their fits:
\begin{eqnarray}
\mbox{LHCb:} \; \; \; \; \tau_{B_s} & = & 1.520 \pm 0.020 \; \mbox{ps} \; .
\\  
\mbox{CDF:} \; \; \; \;  \tau_{B_s} & = & 1.528 \pm 0.021 \; \mbox{ps} \; .
\\ 
\mbox{D0:} \; \; \; \;   \tau_{B_s} & = & 1.443^{+0.038}_{-0.035} \; \mbox{ps} \; .
\end{eqnarray}
Taking the most precise value \cite{LHCb:2012} and the most recent standard model 
determination \cite{Lenz:2011ti}
we find an impressive agreement
        \begin{eqnarray}
\frac{\tau_{B_s}}{ \tau_{B_d}}^{\rm Exp} = 1.001 \pm 0.014\; , &&
\frac{\tau_{B_s}}{ \tau_{B_d}}^{\rm SM} = 0.996 ... 1.000 \; .
        \end{eqnarray}
This again confirms the framework of the HQE. For a more sophisticated comparison
of course all lifetime measurements have to be combined properly - this was done
also recently by HFAG \cite{Asner:2010qj} (for this average the most recent LHCb number\cite{LHCb:2012}
is not yet included, but the preceding number from \cite{LHCb:2011aa}):
\begin{equation}
\frac{\tau_{B_s}}{\tau_{B_d}} = 0.984 \pm 0.011 \; .
\end{equation}
This new average agrees within about $1 \sigma$ with the very precise theoretical prediction and the HQE seems
to work with a high accuracy.
\\
The new lifetime values for the $B_s$ meson can also be used to update the theory predictions
for the effective lifetimes.
Previously \cite{Fleischer:2011cw} $\tau_{B_s} = 1.477^{+0.021}_{-0.022}$ ps was used.
Replacing this number by the most recent LHCb measurement\cite{LHCb:2012}  one finds
the new SM model predictions for the effective lifetimes and the flavour-specific lifetime listed below.
This can be compared with the experimental numbers for $\tau^{\rm Eff} (K^+ K^-) $
from LHCb \cite{LHCb:2012b}, for $\tau^{\rm Eff} (\psi f_0)$ from CDF \cite{Aaltonen:2011nk} and the
average of the flavour specific lifetime from HFAG \cite{Asner:2010qj}:
        \begin{displaymath}
        \begin{array}{|l||c|c|c|}
        \hline
                                 & \mbox{Exp.} &  \mbox{SM-old} & \mbox{SM-new}
        \\
        \hline
        \hline
        \tau^{\rm Eff} (K^+ K^-) \; \mbox{in ps}& 1.468 \pm 0.046 & 1.390 \pm 0.032 & 1.43 \pm 0.03
        \\
        \hline
        
        \tau^{\rm Eff} (\psi f_0)\; \mbox{in ps}& 1.70 \pm 0.12 & 1.582 \pm 0.036& 1.63 \pm 0.03
        \\
        \hline
        \tau^{\rm FS}      \; \mbox{in ps}      & 1.463   \pm 0.032 &    ---         & 1.54 \pm 0.03
        \\
        \hline
        \end{array}
        \end{displaymath}
For the SM predictions of the effective lifetimes the dominant sources of the error are
the theory value of $\Delta \Gamma_s$ (about $ \pm 0.02$)
and the experimental values for $\tau_{B_s}$.
The effective lifetimes agree within $1 \sigma$, while the experimental value for the 
flavour-specific lifetime is about $2.4 \sigma$ below the theoretical central value. Here also new, more precise data will be very desirable.

\subsection{First measurement of $\Delta \Gamma_s$}

In Moriond LHCb presented the first measurement ($>5 \sigma$) of a non-zero value of the decay rate difference
in the neutral $B_s$ system \cite{LHCb:2012}:
       \begin{eqnarray}
           \Delta \Gamma_s & = &
                          \left( 0.116 \pm 0.019 \right) \mbox{ps}^{-1} \; .
        \end{eqnarray}
Naively this corresponds to a statistical significance of $6.1 \sigma$. 
$\Delta \Gamma_s$ was obtained from
the angular analysis\cite{Dunietz:2000cr} in the decay $B_s \to J/\psi \phi$.
CDF\cite{CDF:2012}  and D0\cite{Abazov:2011ry}  also performed similar analyses to get:
\begin{equation}
\begin{array}{ll}
\mbox{D0 8fb}^{-1}   &  \Delta \Gamma_s = \left( 0.163 \pm 0.065 \right) \mbox{ps}^{-1} \; ,
\\
\mbox{CDF 9.6fb}^{-1} & \Delta \Gamma_s = \left(  0.068  \pm 0.026 \pm  0.007 \right) \mbox{ps}^{-1} \; .
\end{array}
\end{equation}
The new HFAG average \cite{Asner:2010qj} reads (for this average the most recent LHCb number\cite{LHCb:2012}
is not yet included, but the preceding number from \cite{LHCb:2011aa}):
       \begin{eqnarray}
           \Delta \Gamma_s & = &
                          \left( 0.100 \pm 0.013 \right) \mbox{ps}^{-1} \; .
        \end{eqnarray}
There is also a long history of theoretical predictions for $\Delta \Gamma_s$. The first estimates stem from
1977 \cite{Ellis:1977uk} where $\Delta \Gamma_d / \Gamma_d = 1/6$ was found (for $B_d$ mesons!), 
which is very close to the current value for $\Delta \Gamma_s / \Gamma_s$.
As can be seen from Eq.(\ref{convergence}) NLO-QCD corrections, subleading HQE corrections and 
the impact of lattice parameters for $\Delta \Gamma_s$ are ample. 
Therefore the numerical values of old calculations have to be taken
with a pinch of salt, even if some of them are by accident close to the current experimental number.
\\
The subleading $1/m_b$-corrections were determined in 1996 \cite{Beneke:1996gn}.
The dominant NLO-QCD corrections were calculated in 1998 \cite{Beneke:1998sy}, subdominant ones in 2003 
\cite{Beneke:2003az,Ciuchini:2003ww}. In 2006 \cite{Lenz:2006hd} more theoretical improvements were 
included - now also the relatively well-known bag parameter $B_{B_s}$ gives the dominant contribution.
Before that the dominant contribution came from the bag parameter $\tilde{B}_S$, which describes 
the matrix elements of the operator $\tilde{Q}_S$, see. Eq.(\ref{tildeQS}).
First lattice estimates of $\tilde{B}_S$ appeared from 2000 on 
\cite{Yamada:2000ym,Gimenez:2000en,Becirevic:2000sj}\cite{} 
So, before 2000 in principle no reliable estimates of $\Delta \Gamma_s$ could be given at all, 
because of missing information and before 2006 estimates were affected by large uncertainties.
More recent evaluations \cite{Lenz:2006hd,Lenz:2011ti} read \footnote{The Rome Group will soon 
present \cite{Rome} a numerical update, which yields $\Delta \Gamma_s / \Gamma_s = 0.149 \pm 0.015$.
This corresponds to $ \Delta \Gamma_s = 0.098 \pm 0.010 \; \mbox{ps}^{-1}$ and is in perfect agreement
with the measurement. It also agrees with the result from \cite{Lenz:2011ti}.}
        \begin{eqnarray}
        2006: &&  \Delta \Gamma_s =  
                   { \left( 0.096 \pm 0.036 \right) \mbox{ps}^{-1}} \; ,
        \\
        2011: &&  \Delta \Gamma_s =  
                   { \left( 0.087 \pm 0.021 \right) \mbox{ps}^{-1}} \; .
        \label{DGsSM}
        \end{eqnarray}
We will stick to Eq.(\ref{DGsSM}) as our theory prediction for $\Delta \Gamma_s$. 
Comparing experiment with theory we again find a perfect agreement:
\begin{equation}
\frac{\Delta \Gamma_s^{\rm Exp.}}{\Delta \Gamma_s^{\rm SM}}
= \frac{0.100 \pm  0.013}{0.087 \pm 0.021} = 1.15 \pm 0.32 \; .
\label{proof}
\end{equation}
This result proofs experimentally that the HQE works also for the calculation of $\Gamma_{12}^q$.
The issue of the proper calculation of $\Gamma_{12}^q$ gave raise to numerous discussions 
in the literature, see e.g. \cite{Bigi:2001ys} and references and citations of this paper. This 
issue is now settled.
\\
Now several comments to Eq.(\ref{proof}) are appropriate:
\begin{itemize}
\item Eq.(\ref{proof}) shows that the HQE works for $\Gamma_{12}^q$ with an accuracy of
      $15 \% \pm 32 \%$. Therefore the question is not anymore, whether the HQE is appropriate for 
      $\Gamma_{12}^q$, but how precise is the HQE. Does it work to an accuracy of $30\%$? Or maybe even to
      an accuracy of $10\%$?
      This can be answered by further reducing the theoretical and the experimental error.
\item The uncertainty in Eq.(\ref{proof}) is dominated by the theoretical error, but an improvement of the
      experimental error will also be helpful. Here new data from LHCb and the two planned 
      Super-B-factories \cite{O'Leary:2010af,Bona:2007qt,Abe:2010sj} will be very important.
\item The theory prediction for $\Gamma_{12}^q$ has again a crucial dependence on non-perturbative 
      parameters. In particular it is also proportional to $f_{B_s}^2$.
      For the theory prediction we have used again $f_{B_s} = 231 \pm 15$ MeV.
      As in the case of $M_{12}^q$ we also show the results, if new evaluations of the decay constant
      are used \cite{McNeile:2011ng,Bazavov:2011aa}.
      \begin{equation}
      \begin{array}{l|c|c}
      \mbox{Reference} & f_{B_s} & \Delta \Gamma_s
      \\
      \hline \hline
      \mbox{HPQCD (1110.4510)}          & 225 \pm  4 \; \mbox{MeV} & 0.083 \pm 0.017 \; \mbox{ps}^{-1}
      \\
      \mbox{Fermilab/MILC (1112.3051)} & 242 \pm 10 \; \mbox{MeV} & 0.096 \pm 0.021 \; \mbox{ps}^{-1}
      \end{array}
      \label{newDeltaGamma}
      \end{equation}
      In contrast to $M_{12}^q$ now the error is not reduced so strongly, if one uses smaller errors for the
      decays constant.
      For the same reasons as discussed in relation to $\Delta M_s$ we will use the prediction from
      Eq.(\ref{DGsSM}).
\item To see, what is necessary to increase the theoretical accuracy, we compare the error budget
      from the theory prediction of $\Delta \Gamma_s$ in 2006 \cite{Lenz:2006hd} with the 
      one from 2011\cite{Lenz:2011ti}:
\begin{displaymath}
\begin{array}{|c||c|c|}
\hline
\Delta \Gamma_s^{\rm SM} &  2011  & 2006
\\
\hline
\hline
\mbox{Central Value}   &  0.087 \, \mbox{ps}^{-1} &  0.096 \, \mbox{ps}^{-1}
\\
\hline
\delta ({\cal B}_{\widetilde R_2})       &   { 17.2 \% }               &  {15.7 \%}
\\
\hline
\delta (f_{B_s})       &  {13.2 \%}                 &  {33.4 \%} 
\\
\hline
\delta (\mu)           &   7.8 \%                 &  13.7 \%
\\
\hline
\delta (\widetilde{{\cal B}}_{S,B_s})           &   4.8 \%                 &  3.1 \%
\\
\hline
\delta ({\cal B}_{R_0})       &    3.4 \%                &   3.0 \%
\\
\hline
\delta (V_{cb})        &    3.4 \%                &  4.9 \%
\\
\hline
\delta ({\cal B}_{B_s})             &   2.7 \%                 &  6.6 \%
\\
\hline
...                            &   ....                & ...
\\
\hline
\hline
\sum \delta            &  24.5 \%                & 40.5 \%
\\
\hline
\end{array}
\end{displaymath}
      Currently the dominant theoretical uncertainty stems from the power suppressed operator
      $R_2$. These matrix elements have currently only been estimated with vacuum insertion
      approximation. Numerical subleading effects within the QCD sum rule approach have been
      determined in \cite{Mannel:2011zza,Mannel:2007am}. Here a more precise determination
      of $B_{R_2}$ is mandatory in order to improve the theoretical accuracy of $\Gamma_{12}^q$.
\item For a long time the dominant error source of the theory prediction of $\Delta \Gamma_s$
      was $f_{B_s}$, currently the dependence on $f_{B_s}^2$ is the second largest error.
      Thus one might want to get rid off the dependence on $f_{B_s}$ by considering
      the ratio $\Delta \Gamma_s / \Delta M_s$ - here one implicitly also assumes that no new physics 
      acts in $\Delta M_s$. In this ratio the overall factor $f_{B_s}^2$ cancels, as well as the
      bag parameter $B_{B_s}$ of the operator $Q$, given in Eq.(\ref{Q}):
      \begin{eqnarray}
      \frac{\Delta \Gamma_s}{\Delta M_s}  & = & 10^{-4} \cdot
      \left[ {46.2}  
      + { 10.6 \frac{ \tilde B_S'}{B}}  
      - \left(13.2 \frac{B_{\tilde{R}_2}}{B} - { 2.5 \frac{B_{R_0}}{B}} 
      + 1.2 \frac{B_R}{B}  \right) \right] 
      \nonumber 
      \\
      & = & 0.0050 \pm 0.0010 \; .
      \end{eqnarray}
      The numerical dominant term is now completely free of hadronic uncertainties, the remaining
      dependence on hadronic parameters is always via ratios of bag parameters.
      Comparing this theory prediction \cite{Lenz:2011ti} with the experimental number presented 
      by HFAG  \cite{Asner:2010qj}
      one gets
      \begin{eqnarray}
      \left( \frac{\Delta \Gamma_s}{\Delta M_s} \right)^{\rm Exp}
      \left/ 
      \left( \frac{\Delta \Gamma_s}{\Delta M_s} \right)^{\rm SM} \right.
      & =  & 1.12 \pm 0.27 \; .
      \end{eqnarray}
      Once more an impressive agreement with the HQE prediction and the assumption that no new 
      physics acts in $\Delta M_s$ or $\Delta \Gamma_s$.
\end{itemize}

\subsection{$\Delta \Gamma_s$ from $Br(B_s \to D_s^{(*)+} +D_s^{(*)-})$?}

In the literature one finds regularly the following relation
\begin{equation} 
\frac{\Delta \Gamma_s^{\rm CP}}{\Gamma_s} = 2 Br \left(B_s \to D_s^{(*)+} +D_s^{(*)-} \right)
\label{nottouse}
\end{equation}
and one might of course be tempted to use Eq.(\ref{nottouse}), 
to determine $\Delta \Gamma_s^{\rm CP} = |\Gamma_{12}^s|$ directly from ``simple'' branching ratio measurements.
This equation was derived in 1993\cite{Aleksan:1993qp} to show the equivalence of 
the exclusive and inclusive approach for calculating $\Delta \Gamma_s$. It holds in
the limit {$ m_c \to \infty; m_b - 2m_c \to 0; N_c \to \infty$}. This limit 
corresponds to neglecting the 3-body final state contributions to $\Gamma_{12}^s$. In 
1993 Aleksan et al. estimated\cite{Aleksan:1993qp}
       \begin{equation}
          \frac{\Delta \Gamma_s}{\Gamma_s} \propto {\cal O} (0.15) \; , 
      \label{Aleksan}
      \end{equation}
which is amazingly close to the current experimental value  \cite{Asner:2010qj}.
This analysis was redone recently\cite{Chua:2011er}.
With up to date values of the input parameters one gets: 
the 2-body final states contribute {$0.100 \pm 0.030$} to 
$\Delta \Gamma_s / \Gamma_s$. This number is the updated value of Eq.(\ref{Aleksan}). It is slightly below the
HQE prediction in Eq.(\ref{DGsSM}). Moreover it turned out that the 3-body final state contributions -
which are neglected in the derivation of Eq.(\ref{nottouse}) - are quite sizable, they contribute about 
{0.06...0.08} to $\Delta \Gamma_s / \Gamma_s$. Since this is comparable to the 2-body final states,
the above approximation turns out to be a bad one. We repeat here the suggestion from\cite{Lenz:2006hd}:
\\
{\it We strongly discourage from the inclusion of 
$ Br( Bs \to D^{(*)+} + D^{(*)-})$ in
averages with $\Delta \Gamma_s$  determined from clean methods.}

\subsection{HQE at work: semi leptonic CP asymmetries and mixing phases}

In the previous sections we have shown that the HQE describes inclusive decay of b-hadrons very accurate.
Even the prediction of $\Gamma_{12}^s$ agrees with an accuracy of about $30\%$ with the experiment. To make 
more precise statements, some non-perturbative progress is mandatory.
\\
Because of this success we use now the HQE to 
predict\cite{Lenz:2011ti} also quantities that might be very sensitive to 
new physics effects like semi leptonic CP asymmetries and mixing phases.
        \begin{eqnarray}
        a_{fs}^s =  \left(1.9  \pm 0.3 \right) \cdot 10^{-5} \; ,
        && 
        \phi_s   =  0.22^\circ \pm 0.06^\circ \; ,
        \\
        a_{fs}^d =  - \left(4.1 \pm 0.6 \right) \cdot 10^{-4} \; ,
        && 
        \phi_d  =  {-4.3^\circ} \pm 1.4^\circ \; ,
        \\
        A_{sl}^b = 0.406 a_{sl}^s + 0.594 a_{sl}^d & = & (-2.3 \pm 0.4) \cdot 10^{-4} \; .
        \end{eqnarray}
The semi leptonic CP asymmetries turn out to be extremely small, as well as the mixing phase in the 
$B_s$-system. Therefore any measurement of a sizable value would be a clear indication of new physics.
The so-called di-muon asymmetry $A_{sl}^b$ is a linear combination of the semi-leptonic asymmetries in the
$B_d$ and $B_s$ systems.
As mentioned in the the introduction there was a considerable excitement in the past, because we had
several hints for huge new physics effects in $B$-mixing.
Before the Lepton-Photon conference 2011 a fit\cite{Lenz:2010gu} 
of flavour observables resulted in a huge deviation of the $B_s$ 
mixing phase from the tiny SM prediction:
        \begin{eqnarray}
        \phi_s          & = & \phi_s^{\rm SM} + \phi_s^{\Delta} \; ,
        \\
        \phi_s^\Delta   & = & -51.6^\circ \pm 12^\circ  \; ,
        \end{eqnarray}
where $\phi_s^\Delta$ denotes the deviation from the SM value of the $B_s$-mixing phase.
Moreover the D0 collaboration measured\cite{Abazov:2011yk,Abazov:2010hj,Abazov:2010hv} a large deviation
of the di-muon asymmetry from the theory expectation: 
        \begin{eqnarray}
        A_{sl}^b & = & -(7.87 \pm 1.72 \pm 0.93) \cdot 10^{-3} \; .
        \end{eqnarray}
This number is by a factor of 34 larger than the SM expectation, the statistical significance of the 
deviation is $3.9 \sigma$.
The di-muon asymmetry currently still has its large value, while the hints for large deviations
of $\phi_s$ from zero have disappeared. A fit of the most recent data, e.g.\cite{LHCb:2012} 
from $B_s \to J/\psi \phi$ gives
        \begin{eqnarray}
        \phi_s^\Delta         & = & -0.1^\circ \pm 6.1^\circ \; . 
        \end{eqnarray}
This number is perfectly consistent with the SM, while still some ample deviations are allowed. 
Current experimental bounds on the individual semi leptonic CP asymmetries are given by 
HFAG\cite{Asner:2010qj}
        \begin{eqnarray}
        a_{fs}^s & = & \left(-1050 \pm 640 \right) \cdot 10^{-5} \; ,
        \\
        a_{fs}^d & = & -\left(33 \pm 33 \right) \cdot 10^{-4}    \; .
        \end{eqnarray}
These remaining deviations will be investigated systematically below.

\subsection{The decay rate difference in the $B_d$-system}
\label{DGd}
The SM predicts \cite{Lenz:2011ti} also a tiny width difference in the $B_d$ system.
        \begin{eqnarray}
        \frac{\Delta \Gamma_d}{\Gamma_d} & = & 
         \left(4.2 \pm 0.8 \right) \cdot 10^{-3}  \; .
        \end{eqnarray}
This can be compared to the current most precise bound\cite{Higuchi:2012kx} from Belle.
        \begin{eqnarray}
        \frac{\Delta \Gamma_d}{\Gamma_d} & = & 
         \left(-17 \pm 21\right) \cdot 10^{-3}  \; .
        \end{eqnarray}
$\Delta \Gamma_s$ is governed by the CKM-leading tree-level decay $ b \to c \bar{c} s$. So any sizable
new physics effect to this decays would also affect strongly many other flavour observables.
$\Delta \Gamma_d$ is governed by Cabibbo-suppressed decays, hence the bounds on new physics effects
in $\Delta \Gamma_d$ are weaker and the real value of $\Delta_d$ might deviate from
its almost vanishing SM value. This interesting null-test of the SM was discussed recently 
in \cite{Gershon:2010wx}.

\section{New Physics in $B$-mixing}

\begin{figure}
\includegraphics[width=0.8\textwidth,angle=0]{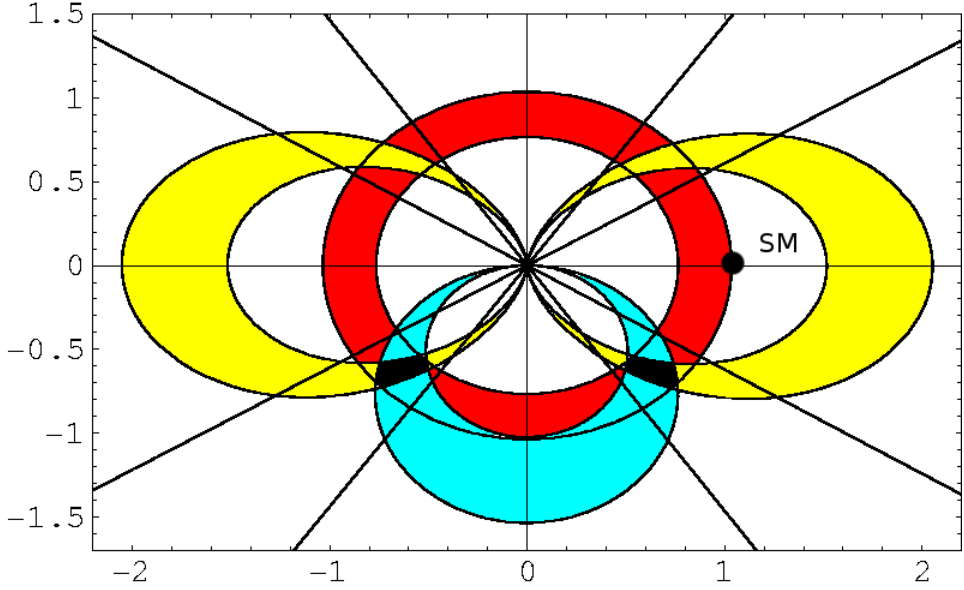}
\caption{Bounds on the complex $\Delta_q$-plane from $\Delta M_q$ (red), $\Delta \Gamma_q / \Delta M_q$
(yellow), $a_{sl}^q$ (turquoise) and direct determinations of $\phi_q$ (black rays).}
\label{Delta}
\end{figure}

New physics in $B$-mixing can be paramterized model independently in the following way
\begin{eqnarray}
\Gamma_{12}^q =  \Gamma_{12}^{q, \rm SM}\, ,
&&
M_{12}^q  =  M_{12}^{q,\rm SM} \cdot { \Delta_q} \, ,
\, \, \, \, \, \, \,  
{ \Delta_q} = 
{|\Delta_q|} e^{i  \phi^\Delta_q} \; .
\label{NP}
\end{eqnarray}
Here we assume that new physics acts only in $M_{12}^q$, while $\Gamma_{12}^q$ is given by the SM
prediction. Although this is expected to hold within the hadronic uncertainties (up to $25 \%$),
we will also investigate new physics effects in $\Gamma_{12}^q$ in the next section. 
\\
Using the parameterization of Eq.(\ref{NP}) one can express the observables in the
mixing system in terms of the SM predictions for $M_{12}^q$ and $\Gamma_{12}^q$ and in terms
of the complex parameter $\Delta_q$:
\begin{eqnarray}
 \Delta M_q  & = & 2 | M_{12}^{q,\rm SM} | \cdot { |\Delta_q |}  \; ,
\label{DMNP}
\\
\Delta \Gamma_q  & = & 2 |\Gamma_{12}^q| 
\cdot \cos \left( \phi_q^{\rm SM} + { \phi^\Delta_q} \right) \; ,
\label{DGNP}
\\
\frac{\Delta \Gamma_q}{\Delta M_q} 
&= &
{ \frac{|\Gamma_{12}^q|}{|M_{12}^{q, \rm SM}|}} 
\cdot \frac{\cos \left( \phi_q^{\rm SM} + {\phi^\Delta_q} \right)}
{ |\Delta_q|} \; ,
\label{DGDMNP}
\\
a_{fs}^q 
&= &
{\frac{|\Gamma_{12}^q|}{|M_{12}^{q,\rm SM}|}} 
\cdot \frac{\sin \left( \phi_q^{\rm SM} + {\phi^\Delta_q} \right)}
{|\Delta_q|} \; .
\label{aslNP}
\end{eqnarray}
Noticing $ \sin (\phi_s^{\rm SM}) \approx 1/240$, one sees that new physics
easily could enhance the semi leptonic asymmetries in the $B_s$-system by a 
factor of almost 250, if a sizable phase $\phi_s^\Delta$ is present in $M_{12}^s$.
 \\
The relations of Eq.(\ref{DMNP}) - Eq.(\ref{aslNP}) can be used to obtain bounds on the 
complex $\Delta_q$-plane. As inputs one needs the SM predictions for $M_{12}^q$ and $\Gamma_{12}^q$, 
as well as the experimental numbers for $\Delta M_q$, $\Delta \Gamma_q$, $a_{sl}^q$ and 
$\phi_q = \phi_q^{\rm SM} + {\phi^\Delta_q}$.
For the hypothetical case ${ |\Delta_q| = 0.9}$ and ${ \phi^\Delta_q = - \pi / 4}$ 
one gets the bounds of Fig.(\ref{Delta}) in the { complex $\Delta_q$-plane}.
Eq.(\ref{DMNP}) gives a bound on the modulus of $\Delta_q$, this corresponds to a circle
in the complex $\Delta_q$-plane (red region in Fig.(\ref{Delta})). Eq.(\ref{DGNP}) gives a direct bound
on the phase, this corresponds to the black rays in  Fig.(\ref{Delta}). Because of the 
relatively large theory uncertainties in $\Gamma_{12}^q$, it is however much more advantageous
to use a direct determination of $\phi_q$, e.g. for $\phi_s$ 
from the angular analysis of decays like $B_s \to J/ \psi \phi$ 
or  $B_s \to J / \psi f_0$\cite{:2012dg}. Eq.(\ref{DGDMNP}) and  Eq.(\ref{aslNP}) constrain simultanously 
the modulus of $\Delta_q$ and the phase $\phi_q^\Delta$. $\Delta \Gamma_q / \Delta M_q$ 
gives the yellow region in Fig.(\ref{Delta}), while $a_{sl}^q$ gives the turquoise one.
\\ 
Combining all data till march 2012 and allowing only for new physics in $B_d$- and $B_s$-mixing 
(i.e. neglecting e.g. new physics penguin contributions), one gets the following bounds
on the complex $\Delta_d$- and $\Delta_s$-planes\cite{Lenz:2012az} 
(update of \cite{Lenz:2010gu}).
\begin{center}
\includegraphics[width=0.49\textwidth,angle=0]{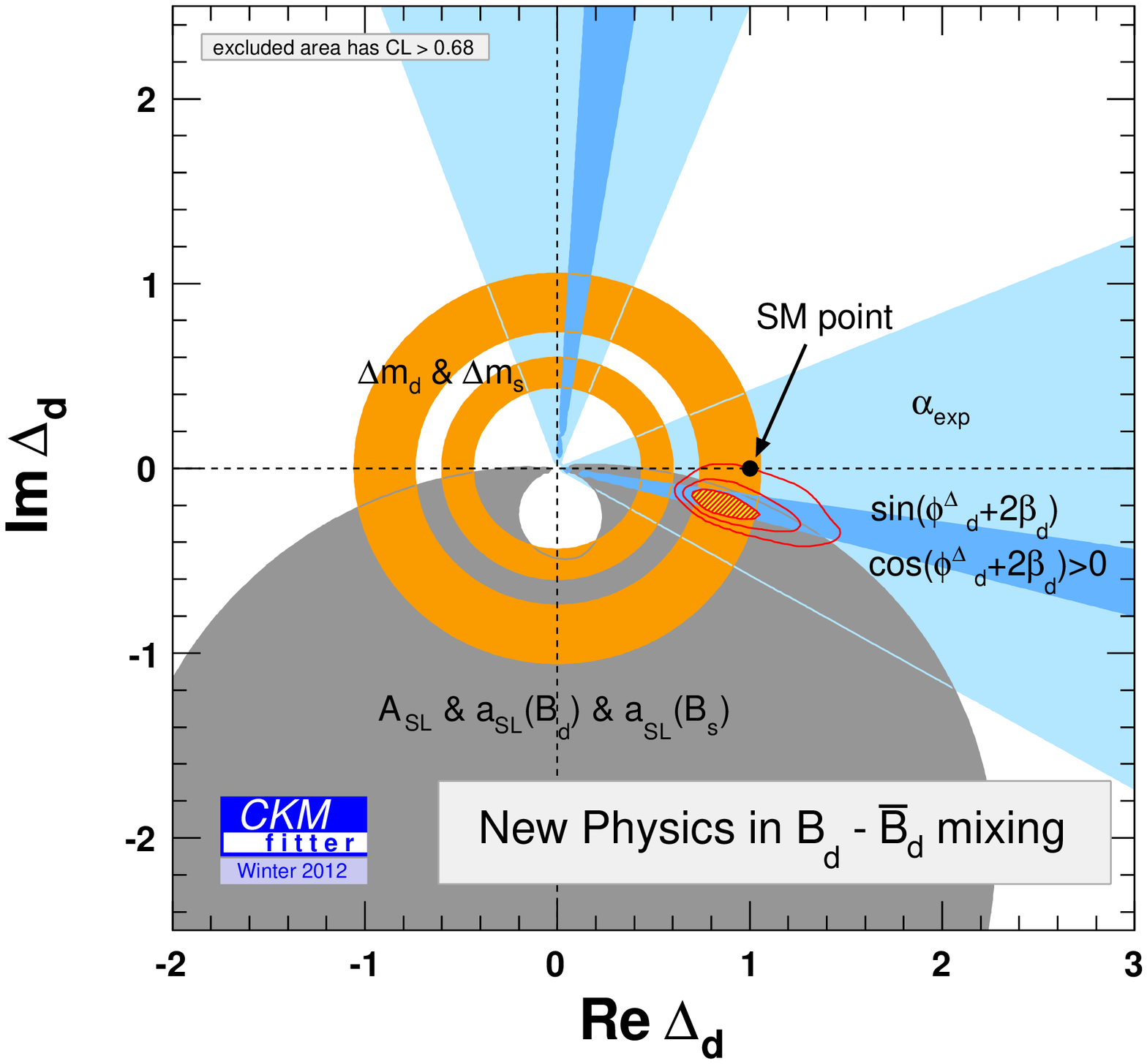}
\hfill
\includegraphics[width=0.49\textwidth,angle=0]{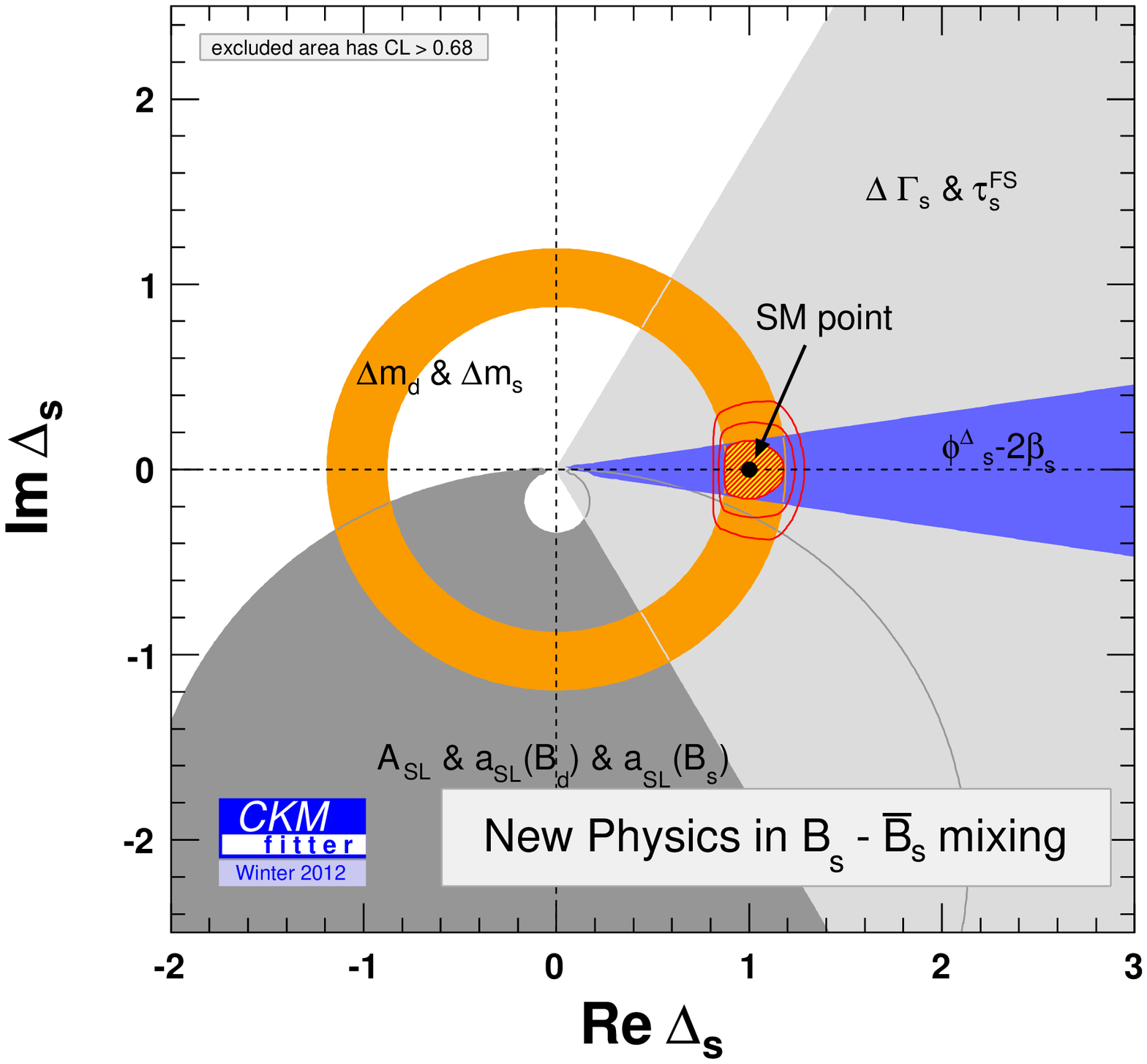}
\end{center}
In 2010\cite{Lenz:2010gu} new physics in $B$-mixing could very well accommodate
the different deviations from the SM expectations, seen at that time. 
This is not the case anymore in 2012\cite{Lenz:2012az}.
There is now a tension between the direct determination of $\phi_s$ and the di-muon asymmetry.
\\
In the $B_d$-system, new physics in $M_{12,d}$ can resolve the discrepancy between
$B \to \tau \nu$ and direct determinations of  $\sin 2 \beta$. In the $B_s$-system
everything looks SM-like although still sizable values for $\phi_s^\Delta$ are possible.
Just recently a second (symmetric) solution in the complex $\Delta_s$-plane was
excluded\cite{Aaij:2012eq}. We also would like to note that in
\cite{Lenz:2012az} no tension is found for $\epsilon_K$.
\\
To improve further the bounds on the complex $\Delta_q$-planes, more precise data are 
necessary.

\section{New Physics in $\Gamma_{12}$}
\begin{figure}
\includegraphics[width=0.8\textwidth,angle=0]{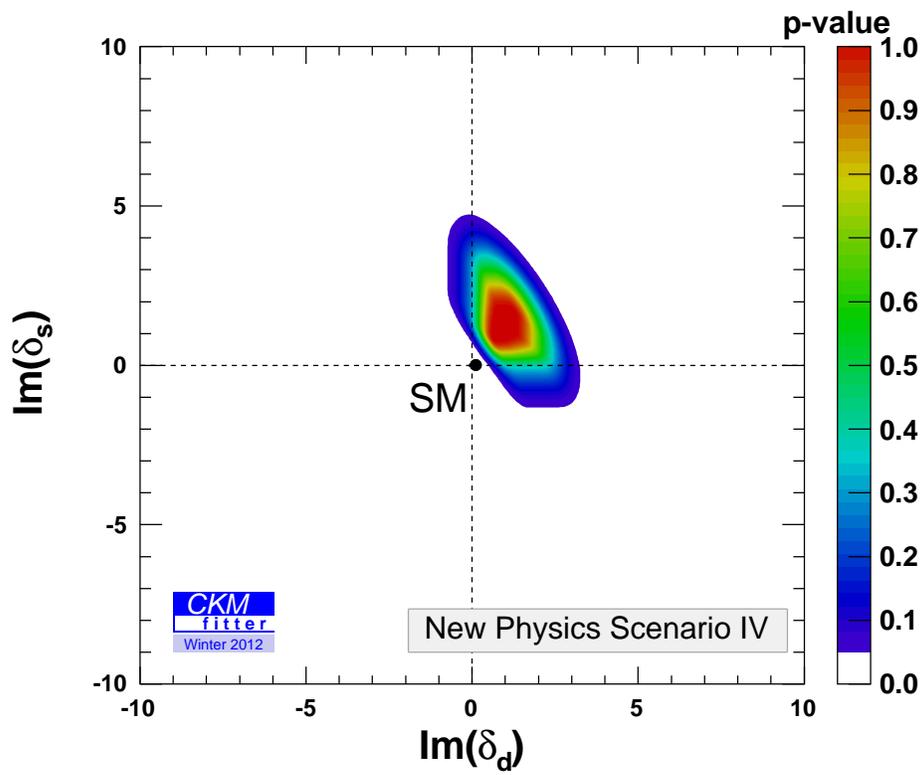}
\caption{Bounds on the parameter Im($\delta_q$), if only $\Delta M_q$, $\Delta \Gamma_q$ and
  $a_{sl}^q$ are taken into account. Large values of Im($\delta_s$) violate however the perfect agreement
  of experiment and theory for quantities
  like $\tau_{B_s} / \tau_{B_d}$,... (see text).}
\label{NPGamma12}
\end{figure}

The theory expression for the di-muon asymmetry can be written in the following way
\begin{eqnarray}
A_{sl}  & = & 
 (0.594 \pm 0.022)(5.4 \pm 1.0) \cdot 10^{-3} \frac{\sin (\phi_d^{SM} + \phi_d^\Delta)}{| \Delta_d |}
\nonumber
\\
&& 
+ (0.406 \pm 0.022)(5.0 \pm  1.1) \cdot 10^{-3} \frac{\sin (\phi_s^{SM} + \phi_s^\Delta)}{| \Delta_s |} \; .
\end{eqnarray}
Since $\Delta_s$ and $\Delta_d$ are bounded from measurements of the
mass differences to be close to one and the sine can be at most one, there exists a 
theoretical upper limit for the di-muon asymmetry.
We use here the fit values of $\Delta_q$ from \cite{Lenz:2012az} to obtain the following upper bounds:
\begin{eqnarray}
A_{sl}  
&  \leq & \left\{ 
\begin{array}{cll}
-1.7  \cdot 10^{-3}:& 1 \sigma \; \mbox{for} \; |\Delta_q|, & 1 \sigma \; \mbox{for} \; \phi_q^\Delta, 
\\
-2.8  \cdot 10^{-3}:& 3 \sigma \; \mbox{for} \; |\Delta_q|, & 3 \sigma \; \mbox{for} \; \phi_q^\Delta, 
\\
-7.5  \cdot 10^{-3}:& 3 \sigma \; \mbox{for} \; |\Delta_q|, & \mbox{set sine to} \; 1 .
\end{array}
\right.
\label{upperbound}
\end{eqnarray}
For the first number the four parameters of $\Delta_q$ (q=s,d) have been chosen to take
the value, which gives the largest di-muon asymmetry, within the allowed $1 \, \sigma$ range of the 
fit in \cite{Lenz:2012az} ,
for the second number, the $3 \, \sigma$ range has been chosen, while for the third number the sine
has been set to one by hand. The last number is purely hypothetical, because such a large value of 
the mixing phase is in contrast to experimental investigations of e.g. $B_s \to J / \psi \phi$
\footnote{This also holds, if one takes into account large new physics penguin contributions
to the decay $b \to c \bar c s$, which could lead to a certain extent to a cancellation between the
penguin phase and $\phi_q^\Delta$. See the discussion in the next section.}.
The above bounds have to be compared with the experimental 
measurement\cite{Abazov:2011yk,Abazov:2010hj,Abazov:2010hv}
\begin{eqnarray}
A_{sl}^{D0} & = &  { \left( -7.8 \pm 2.0 \right) \cdot 10^{-3} } . 
\end{eqnarray}
One finds that the central value of the di-muon asymmetry is larger than {\it theoretically} possible, 
see also \cite{Lenz:2011ww}.
This discrepancy triggered a lot of interest in the literature (see the list of citations for
\cite{Abazov:2011yk,Abazov:2010hj,Abazov:2010hv}).
The theoretical upper limit stems from the fact that we did not allow ample (i.e. larger than the
hadronic uncertainties, which are of the order of $25 \; \%$) deviations of $\Gamma_{12}^q$ from
its SM value. One might of course be tempted to give up this assumption.
If there is no new physics acting in $M_{12}^q$ one would need an enhancement factor for $\Gamma_{12}^q$
of 34 to describe the central value of the di-muon asymmetry, if new physics acts in $M_{12}^q$
one still needs an enhancement factor of about 4.6 (1 $\sigma$-range of Eq.(\ref{upperbound}))
or 2.8 (3 $\sigma$-range of Eq.(\ref{upperbound})).
\\
As discussed above violations of quark hadron duality in the range of $280\% - 3400 \%$ are clearly ruled 
out now, as experiment and theory agree for $\Delta \Gamma_s$ to an accuracy of about $30\%$.
\\
Next one might think of new physics effects in $\Gamma_{12}^q$.
One can modify Eq.(\ref{NP}) to allow for new physics in $\Gamma_{12}^q$:
\begin{eqnarray}
\Gamma_{12}^q =  \Gamma_{12}^{q, \rm SM}    \cdot { \tilde{\Delta}_q} \, ,
&&
M_{12}^q      =  M_{12}^{q, \rm SM}         \cdot { \Delta_q} \, ,
\nonumber \\
\tilde{\Delta}_q = |\tilde{\Delta}_q| e^{i  \phi^{\tilde{\Delta}_q}} \; ,
&&
{ \Delta_q} = {|\Delta_q|} e^{i  \phi^\Delta_q} \; .
\label{NP2}
\end{eqnarray}
Just looking at $\Delta M_q$, $\Delta \Gamma_q$ and $a_{sl}^q$ one
can of course fit any numerical value of these three observables 
with the three parameters $|\Delta_q|$, $|\delta_q|$ and $\phi_q^\Delta-\phi_q^\delta$.
\begin{eqnarray}
 \Delta M_q  & = & 2 | M_{12}^{q, \rm SM} | \cdot { |\Delta_q |}  \; ,
\\
\Delta \Gamma_q  & = & 2 |\Gamma_{12}^{q, \rm SM}|\cdot { |\tilde{\Delta}_q |}  
\cdot \cos \left( \phi_q^{\rm SM} + { \phi^\Delta_q} - \phi^{\tilde{\Delta}_q}\right) \; ,
\\
a_{fs}^q 
&= &
{\frac{|\Gamma_{12}^{q, \rm SM}|}{|M_{12}^{q, \rm SM}|}} 
\cdot \frac{ |\tilde{\Delta}_q|}{|\Delta_q|}
\cdot \sin \left( \phi_q^{\rm SM} + {\phi^\Delta_q}- \phi^{\tilde{\Delta}_q} \right) \; .
\end{eqnarray}
So the situation $\Delta M_q = \Delta M_q^{\rm SM}$, $\Delta \Gamma_q = \Delta \Gamma_q^{\rm SM}$
and $a_{sl}^q = 34 a_{sl}^{q, \rm SM}$ can be described by $|\Delta_q| = 1 $,
$ |\tilde{\Delta}_q |  \cdot \cos \left( \phi_q^{\rm SM} + { \phi^\Delta_q} - \phi^{\tilde{\Delta}_q}\right) = 1$ 
and
$ |\tilde{\Delta}_q|/|\Delta_q| \cdot \sin \left( \phi_q^{\rm SM} + {\phi^\Delta_q}- \phi^{\tilde{\Delta}_q} \right) = 1$, 
which is equivalent to $|\Delta| = 1$, $|\tilde{\Delta}| =34.0147$ and 
$\cos \left( \phi_q^{\rm SM} + {\phi^\Delta_q}- \phi^{\tilde{\Delta}_q} \right) = 0.0293991$.
In this spirit bounds on new physics contributions to $\Gamma_{12}^d$ and $\Gamma_{12}^s$
were obtained in \cite{Lenz:2012az}, see Fig.(\ref{NPGamma12}).
It turned out to be useful to introduce a new parameter $\delta_q$:
\begin{eqnarray}
\delta_q & = & \frac{\frac{\Gamma_{12}^q          }{M_{12}^q          }}
                    {{\rm Re} \left(
                      \frac{\Gamma_{12}^{q, \rm SM}}{M_{12}^{q, \rm SM}}
                              \right)} \; ,
\\
{\rm Re} \left( \delta_q \right) & = & 
                                       \frac{\Delta \Gamma_q}{\Delta M_q}
                                       \left/ \left(
                                       \frac{\Delta \Gamma_q^{\rm SM}}{\Delta M_q^{\rm SM}}
                                       \right) \right. \; ,
\\
{\rm Im} \left( \delta_q \right) & = & 
                                       - a_{sl}^q
                                       \left/ \left( 
                                       \frac{\Delta \Gamma_q^{\rm SM}}{\Delta M_q^{\rm SM}}
                                       \right) \right. \; .
\end{eqnarray}           
The experimental constraint on $\Delta \Gamma_s$ can be fulfilled by Re $\delta_s \approx 1$ and 
the semi leptonic asymmetries only affect Im  $\delta_s$. The real part of $\delta_d$ is almost 
unconstrained: Re $\delta_d \approx -4.3 \pm 5.4$ (see Section \ref{DGd}), 
while the the semi leptonic asymmetries again
affect only the imaginary part of $\delta_d$.
\\
$\Gamma_{12}^s$ is dominated by the Cabibbo-favoured decay $b \to c \bar{c} s$. Thus any sizable
contribution to $\Gamma_{12}^s$ will affect many other observables dramatically. As explained
in \cite{Lenz:2012az} these effects on other observables have to be taken into account, when using
the results of Fig(\ref{NPGamma12})!
Any new physics contribution to $\Gamma_{12}^s$ corresponds to a transition $b \bar s \to X$,
where $X$ can be one or more (SM or new physics) particles and 
$M_X \leq M_{B_s}$. The new operator $(b \bar s  X)$ gives now rise to many different new contributions.
Via diagrams like (if $X$ corresponds to two particles)

\centerline{\includegraphics[width=0.2\textwidth,angle=270]{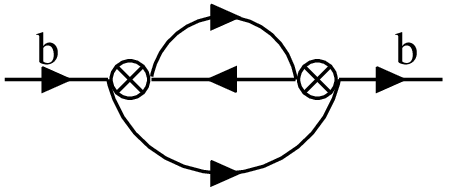}}

\noindent
the $(b \bar s  X)$-operator contributes to the leading term $\Gamma_0$ in the HQE, see. Eq.(\ref{HQEall}).
This affects the total decay rate $\Gamma_s$ and therefore quantities like the semi leptonic branching
ratio, or the branching ratio of a $b$-quark into two, one or zero charm quarks (see the discussion
of the missing charm puzzle in Section(\ref{missing})).
Via diagrams like

\centerline{\includegraphics[width=0.2\textwidth,angle=270]{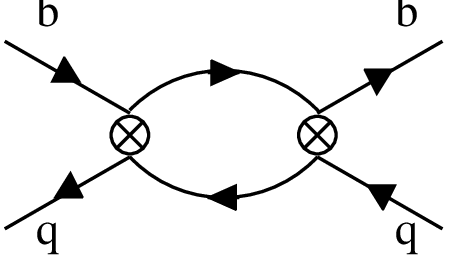}}

\noindent
the $(b \bar s  X)$-operator contributes to the third term $\Gamma_3$ in the HQE, see. Eq.(\ref{HQEall}).
This affects the lifetime ratio $\tau (B_s)/\tau (B_d)$, but also 
the decay rate difference $\Delta \Gamma_s$ and the semi leptonic CP-asymmetries\footnote{Strictly speaking
the contributions to $\Gamma_{12}^q$ and $\Gamma_3$ are of course not equal, see e.g. the discussion in
\cite{Lenz:2008xt}.}. 
It will also give rise to new contributions to $M_{12}^s$.
Via mixing such a new operator can also modify  well constrained quantities like the rare decay 
$b \to s \gamma$.
\\
All in all, any new physics contribution to $\Gamma_{12}^s$ will face severe constraints from many
well measured quantities. Hence it seems impossible that ample (i.e. larger than the hadronic
uncertainties, which are of the order of $25 \%$) new physics effects are present in $\Gamma_{12}^s$.
Unfortunately this statement cannot be proven model independently. At least the operator structure
of the new contribution has to be specified, so that one can also calculate the new contribution to e.g.
$\Gamma_0$ and $\Gamma_3$.
Such an analysis \cite{Bobeth:2011st} has been done for the most promising (because least constrained)
candidate, $X = \tau^+ + \tau^-$ and it turned out that new physics contributions to $\Gamma_{12}^s$
have to be smaller than $30\%$ or $40\%$, depending on the Dirac-structure of the $bs \tau \tau$-operator.
This clearly excludes the possibility of having new physics contributions to
$\Gamma_{12}^s$ in the range of $280\% - 3400 \%$. 
More precise data on $\tau_{B_s}/\tau_{B_d}$, $B_{sl}$, $n_c$, $Br(B_s \to \tau^+ \tau^-)$,... will
further shrink the bound on new physics effects in $\Gamma_{12}^s$ (or give hints for small new physics effects).

Finally there is also the possibility that D0 simply saw a 2.5 $\sigma$ upward 
fluctuation of the result for the di-muon asymmetry and that the actual value 
is below -2.8 per mille, which still would be a large deviation from the SM. 
To get a final clue about that, independent measurements of the semi leptonic 
asymmetries are urgently needed.

\section{Penguin pollution}

In the last section I will discuss  the question to what extent does the value of the di-muon
asymmetry contradict the result of analyses of $B_s \to J/\psi \phi$ from LHCb and Tevatron. 
The angular analysis of the decay $B_s \to J/\psi \phi$ at CDF, D0 and LHCb gives among others
the quantity $S_{\psi \phi}$. If one neglects certain penguin contributions, then  $S_{\psi \phi}$ 
is given in the SM simply by CKM-elements. Its SM expectation reads\cite{Lenz:2011ti}
       \begin{equation}
       S_{\psi \phi}^{\rm SM} = 
       0.036 \pm 0.002 \; .
       \end{equation}
In the presence of new physics $S_{\psi \phi}^{\rm SM}$ will be modified by the new mixing phase
$\phi_s^\Delta$, which also changes the values of the semi leptonic CP-asymmetries 
       \begin{equation}
        S_{\psi \phi}^{\rm SM}  
        \to \sin \left(2 \beta_s {- \phi_s^\Delta} \right)
        = 0.00 \pm 0.11 \; .
        \label{spsiphi}
       \end{equation}
As already mentioned above, in deriving e.g. Eq.(\ref{spsiphi}) certain 
penguin contributions have been neglected.
Since the current experimental bound on  $S_{\psi \phi}$(taken from \cite{Lenz:2012az}) 
has already a small error, this approximation might not be justified any more. It was 
already pointed out in \cite{Faller:2008gt} that a measurement of a relatively small
value of the mixing phase requires a much more stringent control of the penguin contributions.
Including  penguins one gets the general relation
       \begin{equation}
        S_{\psi \phi}^{\rm SM}  
        \to \sin \left(2 \beta_s {- \phi_s^\Delta} 
                            - \delta_s^{\rm Peng, SM}
                            - { \delta_s^{\rm Peng, NP}} \right) \; .
       \label{spsiphigeneral}
       \end{equation}
Looking at Eq.(\ref{spsiphigeneral}) one sees that there is the possibility that
a non-vanishing value of the new mixing phase $\phi_s^\Delta$ is compensated by
SM and new physics penguins. Now it depends on the possible size of the penguins,
to what extent such a cancellation can occur.
\\
 To exactly determine the size of the penguin contributions one would have to solve all
 the non-perturbative dynamics related to this decay, which is clearly not possible.
 For a rough numerical estimate we start with the amplitude
\begin{equation}
A \left( B_{s} \longrightarrow f \right)
=
\langle B_{s} | {\cal H}_{eff} | f \rangle \; ,
\end{equation}
with the effective SM hamiltonian for $b \to c \bar c s$ transitions
\begin{equation}
 {\cal H}_{eff.} = \frac{G_F}{\sqrt{2}}
\left[  \lambda_u \left( C_1 Q_1^u + C_2 Q_2^u \right)
      + \lambda_c \left( C_1 Q_1^c + C_2 Q_2^c \right)
      + \lambda_t \sum \limits_{i=3}^6 C_i Q_i
\right] \;.
\end{equation}
The CKM structure is given by $\lambda_x := V_{xb}^* V_{xs}$;  the decay $b \to c \bar c s$
proceeds via the current-current operators $Q_1^c, Q_2^c$ and the QCD penguin operators $Q_3,...,Q_6$.
$C_1, ... , C_6$ are the corresponding Wilson coefficients.
When the current-current operators $Q_1^u, Q_2^u$ are inserted in a penguin diagram in the effective
theory, they also contribute to $b \to c \bar c s$.
Electro-weak penguins are neglected.
\\
Therefore we have the following structure of the amplitude
 $A \left( B_{s} \longrightarrow f \right)$ 
\begin{eqnarray}
A = \frac{G_F}{\sqrt{2}} & & \left[ 
  \lambda_u    \sum \limits_{i = 1,2} C_i  \langle Q_i^u \rangle^P 
+ \lambda_c    \sum \limits_{i = 1,2} C_i  \langle Q_i^c \rangle^{T+P} 
+ \lambda_t    \sum \limits_{i = 3}^6 C_i  \langle Q_i^u \rangle^T 
\right] \; .
\end{eqnarray}
$\langle Q \rangle^T$ denotes the tree-level insertion of the local operator $Q$, 
$\langle Q \rangle^P$ denotes the insertion of the operator $Q$ in a penguin diagram.
Using unitarity $\lambda_t = - \lambda_u - \lambda_c$ 
we can rewrite the amplitude in the following form.
\begin{eqnarray}
A = \frac{G_F}{\sqrt{2}} \lambda_c  &&
\left[
  \sum \limits_{i = 1,2} C_i \langle Q_i^c \rangle^{T+P} 
- \sum \limits_{i = 3}^6 C_i \langle Q_i^u \rangle^T 
+
\frac{\lambda_u}{\lambda_c} \left( 
  \sum \limits_{i = 1,2} C_i  \langle Q_i^u \rangle^P 
- \sum \limits_{i = 3}^6 C_i \langle Q_i^u \rangle^T \right)
\right] \; .
\end{eqnarray}
The first term gives rise to $\beta_s$ in the
CP asymmetry. The second term (proportional to $\lambda_u$), corresponds to the SM penguin pollution.
To get an idea about the possible size of the second term we notice \cite{CKMfits}
that Re $(\lambda_u   / \lambda_c)< 0.01$ and Im $(\lambda_u   / \lambda_c) < 0.02$.
Moreover the penguin Wilson coefficients $|C_{3,...,6}|$ are typically smaller than 0.04, therefore
one can neglect them in comparison to the Wilson coefficients $C_1, C_2 \approx 1$.
For inclusive decays it turned out that 
\cite{nc_theory} $\langle Q \rangle^P \leq 0.05 \langle Q \rangle^T$,
which would result in a penguin pollution smaller than $0.0005 + 0.0010 i$, i.e. smaller than one 
per mille compared to the contribution of the golden plate mode and thus completely negligible, even if 
one takes into account the current experimental precision.
Of course this estimate is very naive and it might change considerably if the matrix elements of the
exclusive final states are taken into account, but also a non-perturbative enhancement of this estimate by 
a factor of 10 would not change the conclusion.
\\
Here clearly more theoretical work has to be done to quantify the size of penguin contributions,
Experimental strategies to determine the penguin contributions to $B_s \to \psi \phi$ and $B_s \to \psi f_0$
have been discussed in e.g. \cite{Faller:2008gt} and the second reference of \cite{Fleischer:2011cw}. 
An investigation of the decay $B_s -> \psi K_S$ at the LHCb upgrade can also gain important insights
on the size of penguin contributions \cite{Fleischer:1999nz}.
Nevertheless it seems that the penguin pollution in the SM is tiny, except some unknown non-perturbative
would arise.
\\
In principle new physics penguins might be much larger, but it is hard to imagine that a phase 
$\phi_s^\Delta$ of order one can be compensated by penguins, without violation other constraints. 
Therefore the angular analysis of
$B_s \to \psi \phi$ seems  to be in a slight contraction to the central value of the di-muon asymmetry.
\\
Finally we will show that even small penguin contributions have an observable effect and that they
also can be analysed quite model independent \cite{Lenz:2011zz}.
The mass difference, the decay rate difference, the semi leptonic CP asymmetries and $S_{\psi \phi}$
are related via
        \begin{eqnarray}
a_{sl}^s & = & 
- \frac{\Delta \Gamma}{\Delta M} \frac{S_{\psi \phi}}{\sqrt{1- S_{\psi \phi^2}}} {\cdot \delta}
        \label{relateall} \; ,
        \end{eqnarray}
with
\begin{eqnarray}
{\delta} & = &\frac{\tan \left({    \phi_s^{\rm SM} }+ \phi_s^\Delta \right)}
                  {\tan \left({ -2 \beta_s^{\rm SM}}+ \phi_s^\Delta + \delta_s^{\rm peng, SM}+ \delta_s^{\rm peng, NP} \right)} \; .
\end{eqnarray}
All SM phases are small:
$\phi_s^{\rm SM} = 0.22^\circ \pm 0.06^\circ$ and $- 2 \beta_s = \left(2.1 \pm 0.1 \right)^\circ $.
Nevertheless they are non-negligible\cite{Lenz:2011zz}. This can be seen by drawing $\delta$ 
in dependence of $\phi^\Delta$ for different values of the penguin contributions,
$\delta_s^{\rm peng, SM} + \delta_s^{\rm peng, NP} = 0^\circ, 2^\circ, 5^\circ, 10^\circ$.
\begin{center}
\includegraphics[width=0.9\textwidth,angle=0]{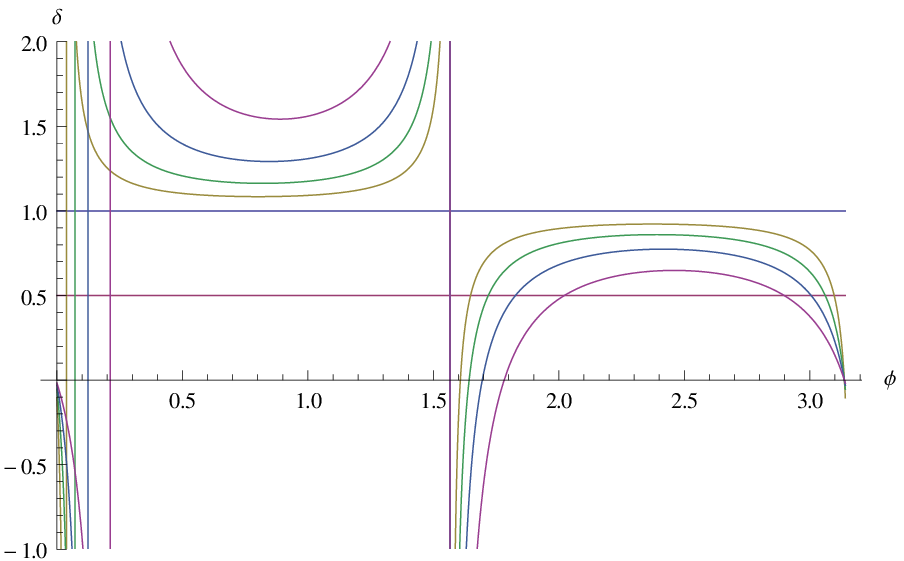}
\end{center}
The curve closest to $\delta = 1$ corresponds to 
$\delta_s^{\rm peng, SM} + \delta_s^{\rm peng, NP} = 0^\circ$, the next one to $ 2^\circ$ and so on.
Because of this huge sensitivity to penguin effects Eq.(\ref{relateall}) can be used to gain some insight
in the size of penguin contributions.

\section{Conclusion}

In Moriond 2012 the first measurement of $\Delta \Gamma_s$ was presented from LHCb.
Combining the LHCb data with the corresponding ones from CDF and D0 and comparing their
average with theory predictions one finds an impressive agreement:
\begin{equation}
\frac{\Delta \Gamma_s^{\rm Exp.}}{\Delta \Gamma_s^{\rm SM}}
= \frac{0.100 \pm  0.013}{0.087 \pm 0.021} = 1.15 \pm 0.32 \; .
\end{equation}
This represents the first experimental proof that the HQE can also be 
applied to the calculation of $\Gamma_{12}^s$, with an uncertainty 
of about $30 \%$, which is an important theoretical insight, because 
$\Gamma_{12}^s$ is expected to be most sensitive to violations
of quark hadron duality. To some extent this is quite amazing, 
because the energy release in the dominant decays, contributing to $\Delta \Gamma_s$ is not so 
large: $m_{B_s} - 2 m_{D_s} \approx 1.428$ GeV.
Currently the experimental and the theoretical errors for $\Delta \Gamma_s$ are of similar size. To improve
the theoretical accuracy, matrix elements of higher dimensional operators
have to be determined non-perturbatively.
\\
Total decay rates, i.e. lifetimes, are expected to be less sensitive to violations of 
quark-hadron duality.
There are now several precise numbers for $\tau (B_s)$ available from  CDF and LHCb,
which also agree perfectly with theory predictions.
\begin{eqnarray}
\frac{\tau_{B_s}}{ \tau_{B_d}}^{\rm Exp} = 1.001 \pm 0.014\; , &&
\frac{\tau_{B_s}}{ \tau_{B_d}}^{\rm SM} = 0.996 ... 1.000 \; .
\end{eqnarray}
This is again a perfect confirmation of the applicability of HQE.
In the case of the lifetime ratios $\tau_{B^+} / \tau_{B_d}$ and
$\tau_{\Lambda_b} / \tau_{B_d}$ no deviation is currently existing, the
precision of any comparison is however strongly limited by our poor knowledge
of the corresponding hadronic parameters and also by the uncertain experimental
value of $\tau_{\Lambda_b}$.
\\
In accordance with this perfect agreement between experiment and theory, 
the new LHC and Tevatron data show no further hints for new physics in $B_s$-mixing.
A model independent fit of all flavour data is consistent with no
new physics in $B_s$-mixing (although there is still some room for sizable deviations 
from the SM expectations) and some small deviations in $B_d$-mixing.
To investigate these issues further a better control over penguin contributions is
mandatory. 
\\
The large central value of the di-muon asymmetry is still an unsolved problem, because
it can not be explained by new physics contributions to $M_{12}^q$ alone, instead one needs
in addition an enhancement of $\Gamma_{12}^s$ compared to its SM value by a factor of $2.8$ up to $34$.
In the text is was discussed, however that $\Gamma_{12}^s$ can not deviate more than about $30\%$
from its SM value (neither due to new physics nor due to violations of quark hadron duality).
On the other hand one should keep in mind that the central value measured by D0 is only about 2.5 $\sigma$
above the theoretical limit. To settle this problem clearly an independent measurement of semi leptonic CP
asymmetries might shed light into the dark.
\\
The success of the HQE for predicting inclusive processes in the $b$-system, raises of course again the
question of the applicability of the HQE to the charm system. Which seems now at first sight to be more 
promising, because for  $D$-mixing the energy release is not so different from $\Delta \Gamma_s$:
$m_D  - 2 m_K  \approx  0.9$ GeV and $m_D  - 2 m_\pi \approx 1.6 $ GeV. 
Of course the situation is much more subtle
because of the huge GIM cancellations in $D$-mixing, see e.g. \cite{D-mix}.

\begin{center}
\includegraphics[width=0.5\textwidth,angle=0]{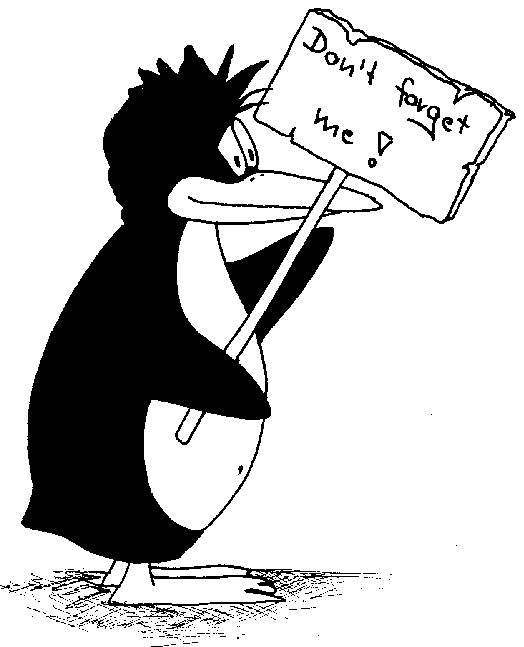}
\end{center}

\section*{Acknowledgements}
I would like to thank the organizers for the invitation; 
Guennadi Borissov, 
Peter Clarke,
Christine Davies, 
Abdelhak Djouadi,
Uli Haisch,
Stephanie Hansmann-Menzemer, 
Andreas Kronfeld,
Giovanni Punzi, 
and
Jonas Rademaker
for numerous discussions in the bar (which resulted also in one paper\cite{Djouadi:2012ae}), 
on the ski slope 
and also during preparing the talk and the proceedings; 
Markus Bobrowski, 
Andrzej Buras, 
Otto Eberhardt, 
Robert Fleischer, 
Vladimir Gligorov,
Uli Nierste 
and
Sheldon Stone for comments on the manuscript
and finally DFG for the support via the HEISENBERG programme.

\end{document}